\documentclass[12pt]{article}
\usepackage{fullpage,epsfig,graphics,amsbsy,amssymb,cancel,bbm,cite}
\usepackage{psfrag,color,slashed}
\usepackage{graphicx}
\usepackage{amsfonts,mathdots,dsfont}
\usepackage{amssymb,amsmath,amscd,mathrsfs}
\usepackage{hyperref}
\usepackage{xcolor}
\usepackage{nicefrac,xfrac,units,comment}
\usepackage{tcolorbox}
\definecolor{mintcream}{rgb}{0.96, 1.0, 0.98}
\definecolor{champagne}{rgb}{0.97, 0.91, 0.81}
\definecolor{bubblegum}{rgb}{0.99, 0.76, 0.8}
\definecolor{mistyrose}{rgb}{1.0, 0.89, 0.88}
\definecolor{cobalt}{rgb}{0.0, 0.28, 0.67}
\usepackage[colorinlistoftodos]{todonotes}

\newcommand{\zb}{\bar{z}}

\newcommand{\tb}{\bar{t}}

\newcommand\e{\epsilon}

\def\L{{\cal L}}

\def\bz{{\bar z}}
\def\half{{1\over 2}}

\newcommand{\bea}{\begin{eqnarray}}
\newcommand{\eea}{\end{eqnarray}}
\newcommand{\be}{\begin{equation}}
\newcommand{\ee}{\end{equation}}
\newcommand{\nn}{\nonumber}
\newcommand{\Tr}{\textrm{Tr}}

\def\d{\partial}
\def\bar{\overline}

\def\b{\bar}

\DeclareMathAlphabet{\mathpzc}{OT1}{pzc}{m}{it}

\usepackage{amsthm}
\usepackage{amsmath}
\usepackage{amsfonts}
\usepackage{amssymb}

\theoremstyle{definition}

\begin{document}

\begin{flushright} \small
UUITP-09/22
\\
ITEP-TH-05/22\\
IITP-TH-04/22\\
MIPT-TH-03/22
 \end{flushright}
\smallskip
\begin{center} \Large
{\bf Localizing non-linear ${\cal N}=(2,2)$ sigma model on $S^2$}
 \\[12mm] \normalsize
{\bf Victor Alekseev${}^{a,b,e}$, Guido Festuccia${}^f$, Victor Mishnyakov${}^{a,b,c,d}$, Nicolai Terziev${}^{b,c}$ and Maxim Zabzine${}^f$} \\
[8mm]
{\small {\it $^a$ MIPT, Dolgoprudny 141701, Russia}}\\
{\small {\it $^b$ NRC "Kurchatov Institute", Moscow 117218, Russia}}\\
{\small {\it $^c$ Lebedev Physics Institute, Moscow 119991, Russia}}\\
{\small {\it $^d$ ITMP, MSU, Moscow  119991, Russia}}\\
{\small {\it $^e$ IITP, Moscow 127994, Russia}}\\
{\small\it ${}^f$Department of Physics and Astronomy, Uppsala University,\\ Box 516, SE-75120 Uppsala, Sweden\\}
\end{center}
\vspace{7mm}
\begin{abstract}
We present a systematic study of ${\cal N}=(2,2)$ supersymmetric non-linear sigma models on $S^2$ with the target being a K\"ahler manifold. We discuss their reformulation in terms of cohomological field theory. In the cohomological formulation we use a novel version of 2D self-duality which involves a $U(1)$ action on $S^2$. In addition to the generic model we discuss the theory with target space equivariance corresponding to a supersymmetric sigma model coupled to a non-dynamical supersymmetric background gauge multiplet. 
We discuss the localization locus and perform a one-loop calculation around the constant maps. We argue that the theory can be reduced to some exotic model over the moduli space of holomorphic disks.  
\noindent
\end{abstract}

\eject
\normalsize

\tableofcontents
\section{Introduction}

Starting from the work of Pestun \cite{Pestun:2007rz} numerous exact calculations for supersymmetric gauge theories on spheres and other curved backgrounds in different dimensions have been performed (see \cite{Pestun:2016zxk} 
 for the review of the subject). 
In this work we are interested in two dimensional theories on $S^2$. Supersymmetric ${\cal N}=(2,2)$ gauge theories coupled to matter were constructed on $S^2$ and localization calculations were carried out in  \cite{Benini:2012ui, Doroud:2012xw}. Many ${\cal N}=(2,2)$ two dimensional  non-linear sigma models with Calabi-Yau spaces as target admit a UV description in terms of  ${\cal N}=(2,2)$ gauge theories, the so called the gauged linear sigma models (GLSM), and thus the partition function $Z_{S^2}$ for GLSM should carry the relevant information about the corresponding non-linear sigma model. In \cite{Jockers:2012dk} it has been conjectured that $Z_{S^2}$ for GLSM computes the quantum corrected K\"ahler potential for the K\"ahler moduli space of the corresponding Calabi-Yau. This conjecture led to a prescription on how to deduce the Gromov-Witten invariants from $Z_{S^2}$ for GLSM and some non-trivial checks were performed. Later in \cite{Gomis:2012wy, Gerchkovitz:2014gta, Gomis:2015yaa} arguments were provided to prove this conjecture. For a review of the subject the reader may consult \cite{Morrison:2016bps, Benini:2016qnm} (and references therein). It is also possible to construct GLSM models on $S^2$ that flow to a Calabi-Yau nonlinear sigma model and whose partition function computes the K\"ahler potential for the complex structure moduli space of the Calabi-Yau~\cite{Doroud:2013pka}.
In this work we concentrate on ${\cal N}=(2,2)$ supersymmetric non-linear sigma models on $S^2$ with the target being a K\"ahler manifold, we study these theories in the context of localization. Our main goal is to understand the formulation of these theories and find on which maps they localize. Equivariance with respect to a rotation of the $S^2$ plays a crucial role in the formalism. If the target K\"ahler manifold admits a toric action we also formulate a version of the sigma model with target space equivariance and we study it on $S^2$. 
These models are closely related to GLSM since some of the parameters in a GLSM can be interpreted as the equivariant parameters for the non-linear sigma model in IR. Thus the relation between GLSM and non-linear sigma model is subtle and, as we argue, the non-linear sigma model is not simply a product of an A-model and a $\bar{A}$-model, but contains more information. 
     
Let us briefly sketch the idea behind our construction. The formulation of the standard A-model is based on the notion on 2D self-duality defined on one forms $\Omega^1(\Sigma, X^*(TM))$ with values in $X^*(TM)$. In 2D the Hodge star $\star$ on one forms satisfies $\star^2=-1$ and if we introduce a (almost) complex structure $J$ on $M$ then on $\Omega^1(\Sigma, X^*(TM))$ we can define the operation $J\star$ such that $(J\star)^2=1$. In the A-model the projector $\frac{1}{2}(1+ J\star)$ (or with minus sign) is used to define some fields and the model is localized on the holomorphic maps $\frac{1}{2}(1+ J\star) dX=0$. Our main observation is that on $S^2$ using  the vector field corresponding to the standard $U(1)$ rotation of $S^2$ one can modify the notion of self-duality on $\Omega^1(\Sigma, X^*(TM))$ and, roughly speaking, define a smooth interpolation between $\frac{1}{2}(1+ J\star)$ over north pole of $S^2$ and $\frac{1}{2}(1- J\star)$ over south pole of $S^2$. Our present 2D construction is similar to the  generalization of self-duality on two-forms in 4D suggested in \cite{Festuccia:2018rew, Festuccia:2019akm} in the attempt to explain the original work \cite{Pestun:2007rz}. 
This novel 2D self-duality condition appears naturally from supersymmetry considerations on $S^2$, as we will explain later. Thus using this new self-duality we can formulate a modification of the A-model on $S^2$ and relate it to ${\cal N}=(2,2)$ non-linear supersymmetric sigma models on $S^2$. We present the details of this construction and we discuss the localization of this model. Although this model is formulated on $S^2$, supersymmetry will force the model to localize on holomorphic disks. 

The paper is organized as follows: in section \ref{s:general-sigma} we review some basic facts about ${\cal N}=(2,2)$ non-linear sigma models on $\mathbb{R}^2$ to set some ideas and conventions.  
In section \ref{s:sigma-S2} we go through the construction of ${\cal N}=(2,2)$ non-linear sigma model on $S^2$ with the target being a K\"ahler manifold. We also describe the supersymmetric sigma model coupled to supersymmetric background gauge multiplets when the target space admits some isometries. 
These supersymmetric models can be reformulated in terms of a cohomological field theory as we present in section \ref{s:cohomological}.  We review the standard A-model and explain how  to modify the notion of self-duality for one forms in the context of the sigma model.
We present the cohomological description of the supersymmetric sigma model both without and with target space equivariance.  In section \ref{s:observables} we analyze the observables of the new cohomological theory. Section \ref{s:locus} deals with the localization locus and problems related to the interpretation of the corresponding  PDEs. In this section we suggest the interpretation of the localization locus and discuss  potential problems. 
In section \ref{s:1-loop} we derive the one-loop result around the constant maps, both for the model with only $S^2$ equivariance and for the model with $S^2$ and target space  equivariance. For the case of Calabi-Yau it agrees with the previously conjectured answer. In section \ref{s:summary} we give a summary of the results, we conjecture the full answer and  outline the open problems within our present understanding of the localization calculation on $S^2$. 

\section{\texorpdfstring{${\cal N}=(2,2)$}{N=(2,2)} nonlinear sigma model on \texorpdfstring{$\mathbb{R}^2$}{R2}}\label{s:general-sigma} 

In this section we review standard facts about ${\cal N}=(2,2)$ non-linear sigma models and set the conventions for further discussion. We are interested in two dimensional non-linear sigma models which are defined on the space of maps $X:\Sigma \rightarrow M $ from a two dimensional manifold  $\Sigma$ to a target manifold $M$ with the following action
\be
  S= \int\limits_{\Sigma} dX^\mu \wedge \star dX^\nu g_{\mu\nu}(X)~,
\ee
where $M$ is equipped with a metric $g$ and $\Sigma$ with Hodge star (which on one-forms requires  only a complex structure). The  non-linear sigma model and its different extensions/generalizations play  a prominent role in string theory and in mathematical physics, in particular through their relation to geometry. Here we are interested in the supersymmetric extensions of the non-linear sigma model which are sensitive to a choice of $\Sigma$ equipped with additional geometrical structures. 
  
The standard discussion of ${\cal N}=(1,1)$ and ${\cal N}=(2,2)$ supersymmetric sigma models is performed on $\Sigma = \mathbb{R}^{1,1}$ equipped with flat Lorentz metric. This choice allows to deal with real (Majorana) two-component spinors. The best way to encode the supersymmetric sigma models is to introduce superfields, see \cite{Hitchin:1986ea}\footnote{Although the paper formally deals with 3-dimensional sigma models, all formalism is identical for 2-dimensional models.} for a review of supersymmetric sigma models and their relation to the geometry of the target space $M$. Here we are interested in ${\cal N}=(2,2)$ sigma models which require $M$ to be a K\"ahler manifold \cite{ZUMINO1979203}. A K\"ahler manifold $M$ is equipped with a metric $g$, a closed K\"ahler form $\omega$ and a complex structure $J$ that satisfy: $g_{i\bar{j}} = \partial_i \partial_{\bar{j}} K$,  $\omega = g J$ and  $\omega = i \bar{\partial} \partial K$ where $K$ is a locally defined K\"ahler potential on $M$. Here we use Latin indices $(i,\bar{i})$ to denote complex coordinates. Because the manifold is K\"ahler the form $\omega$ is covariantly constant with respect to the Levi-Civita connection. Hence the Levi-Civita connection coincides with the Chern connection and its only nonzero Christoffel symbols are
\be{}
  \label{Chr}
  \Gamma^i_{j k}= g^{i \bar j}\partial_j g_{k \bar j}~,\qquad \Gamma^{\bar i}_{\bar j \bar k}= g^{j \bar i}\partial_{\bar j} g_{j \bar k}~.  
\ee
In the following we will be interested in manifolds $M$ with isometries which preserve the K\"ahler structure. Let such an isometry be generated by a vector field $k$ (or collection of the vector fields $k_a$ which are commuting in our setting).
Because $k$ preserves the complex structure, that is ${\cal L}_k J=0$, then in complex coordinates 
$$\partial_i k^{\bar{j}}=0~,\qquad \partial_{\bar i} k^j=0~.$$
Because $k$ is a Killing vector for the K\"ahler metric additionally we have that
$$
\partial_i k_{\bar j}+\partial_j k_i=0~. 
$$
Locally this implies the existence of a function $\cal D$ such that 
\be
\label{defDD}
g_{i \bar j} k^i=-i \partial_{\bar j} {\cal D}~,\qquad g_{ j \bar i} k^{\bar i}=i \partial_{ j} {\cal D}~. 
\ee
We assume that the function $\cal D$ also exists globally and we refer to $\cal D$ as Hamiltonian for $k$. 

Introducing the ${\cal N}=(2,2)$ superspace for $\mathbb{R}^{1,1}$ we can write the action as a superspace integral
\be
S= \int d^2 x ~ d^4 \theta~ K(\mathbf{\Phi}, \bar{\mathbf{\Phi}})~,\label{general-flat-action}
\ee
where  $\mathbf{\Phi}$ is a chiral superfield and its complex conjugate $\bar{\mathbf{\Phi}}$ is an anti-chiral superfield.  For detailed conventions and more general sigma models one may consult \cite{Lindstrom:2005zr}. In the next section we discuss the extension of this model to $S^2$ for a general K\"ahler manifold $M$.
 
If the K\"ahler manifold $M$ admits a torus action $\mathbb{T}^k$ preserving the K\"ahler structure then we can discuss another version of supersymmetry. The vector fields $k_a$ corresponding to the torus action generate a global symmetry of the sigma model (\ref{general-flat-action})
\be
  \delta \mathbf{\Phi}^i = \epsilon^a k_a^i (\mathbf{\Phi})~,~~~~\delta \bar{\mathbf{\Phi}}^{\bar{i}} = \epsilon^a k_a^{\bar{i}} (\bar{\mathbf{\Phi}})
\ee
for any constants $\epsilon^a$ since 
\be
  k^i_a \partial_i K(\mathbf{\Phi}, \bar{\mathbf{\Phi}}) +  k_a^{\bar{i}} \partial_{\bar{i}} K(\mathbf{\Phi}, \bar{\mathbf{\Phi}}) = f (\mathbf{\Phi}) + \bar{f}(\bar{\mathbf{\Phi}})~,
\ee
which follows from ${\cal L}_k \omega = {\cal L}_k (i\bar{\partial}\partial K)=0$. The RHS vanishes under integration over superspace. These global symmetries can be gauged by introducing a real vector superfield $\mathbf{V}^a$ and postulating the transformations
\begin{equation}
    \delta \mathbf{\Phi}^i =i \mathbf\Lambda^a k_a^i (\mathbf{\Phi})~,~~~~\delta \bar{\mathbf{\Phi}}^{\bar{i}} = -i \bar{\mathbf \Lambda}^a k_a^{\bar{i}} (\bar{\mathbf{\Phi}})~,~~~~\delta \mathbf{V}^a = i (\bar {\mathbf \Lambda}^a - \mathbf\Lambda^a)
    \label{superfields-gauge}
\end{equation}
where $\mathbf \Lambda^a$ and $\bar{\mathbf \Lambda}^a$ are gauge parameters promoted to chiral and anti-chiral superfields.  There exists a canonical way of constructing a model that is invariant under the gauge transformations~(\ref{superfields-gauge})
\be
   S= \int d^2 x ~ d^4 \theta~ \hat{K}(\mathbf{\Phi}, \bar{\mathbf{\Phi}}, \mathbf{V})~.
\ee
Here we will treat $\mathbf{V}^a$ as a background field specified in Wess-Zumino gauge. The Wess-Zumino gauge does not preserve supersymmetry, but it preserves a combination of supersymmetry and  super-gauge transformations \cite{Wess:1992cp}. We will refer to such combination as the supersymmetry transformation in the background gauge field. This is a nice way to encode target space equivariance into the supersymmetry transformations of the sigma model. 
 
We want to extended these two types of supersymmetric models to $S^2$. For this we need to work in Euclidean signature. Let us discuss first the model on $\mathbb{R}^2$ equipped with the flat Euclidean metric. Now we have to deal with the Weyl spinors ($\mathbb{R}^2$ does not admit the real Majorana spinor representation). We summarize the spinor and superspace conventions in Appendix \ref{sec:notations}.  The superfield action has the form
\be
      S= \int d^2 x ~ d^4 \theta~ K(\mathbf{\Phi}, \tilde{\mathbf{\Phi}})
\ee
where $\mathbf{\Phi}$ is a chiral superfield and $\tilde{\mathbf{\Phi}}$ is an anti-chiral superfield. In Euclidean signature $\mathbf{\Phi}$ and $\tilde{\mathbf{\Phi}}$ are not related by complex conjugation and reality conditions are subtler. In the next section we provide a detailed construction of ${\cal N}=(2,2)$ sigma models on $S^2$, both for general $M$ and for $M$ with isometries coupled to background gauge fields. 

\section{\texorpdfstring{${\cal N}=(2,2)$}{N=2,2} nonlinear sigma model on \texorpdfstring{$S^2$}{S2}}\label{s:sigma-S2}

Here we first review how to place a ${\cal N}=(2,2)$ theory on a round two sphere preserving all four supercharges. This is accomplished by coupling the theory to an appropriate rigid supergravity background. In case the theory possesses an Abelian flavor symmetry we show how to turn on background gauge multiplets preserving supersymmetry. We comment on the possibility of breaking some of the supercharges in order to deform the round sphere while maintaining a $U(1)$ isometry.
Finally we write down the Lagrangian of an ${\cal N}=(2,2)$ non linear sigma model with K\"ahler target space coupled to background supergravity. In case the target space admits an isometry we show how to introduce the coupling to a supersymmetric background gauge multiplet.
 
\subsection{Killing spinors}
 
In order to preserve supersymmetry in curved space we couple the theory to background supergravity. For the case of a ${\cal N}=(2,2)$ theory the supergravity multiplet includes several bosonic fields in addition to the metric~\cite{Closset:2014pda}. These are a connection $a^{(R)}$ for the $U(1)$ R-symmetry, and two auxiliary scalars $\cal H$ and $\widetilde{\cal H}$. Setting to zero the gravitino variation we obtain the Killing spinor equations
 \bea\label{Killing}
 &&(\nabla_m -i a^{(R)}_m) \zeta = -\frac{1}{4} {\cal H}\gamma_m (1-\gamma^3) \zeta -\frac{1}{4} {\widetilde{\cal H}}  \gamma_m(1+\gamma^3) \zeta~,\cr
 &&(\nabla_m +i a^{(R)}_m) \tilde \zeta = -\frac{1}{4} {\cal H} \gamma_m (1+\gamma^3)\tilde \zeta -\frac{1}{2} {\widetilde {\cal H}}\gamma_m (1-\gamma^3)\tilde \zeta~.
 \eea
Each solution to these equations corresponds to a supercharge acting via supersymmetry variations $\delta_\zeta$ and $\delta_{\tilde \zeta}$. For two solutions $\zeta$ and $\tilde \zeta$ we can define the spinor bilinears:
 \be
 \label{bildef}
 v^m= -2\zeta\gamma^m\tilde \zeta~,\qquad s= \tilde \zeta(1-\gamma_3)\zeta~,\qquad \tilde s= \tilde \zeta(1+\gamma_3)\zeta~.
 \ee
On any field\footnote{We set the central charges of $\phi$ to zero. The case with nonzero central charges is considered in~\cite{Closset:2014pda}} $\phi$ of R-charge $r$ the algebra satisfied by the variations $\delta_\zeta$ and $\delta_{\tilde \zeta}$ is
 \bea
 \label{algsimp}
 && \{\delta_\zeta, \delta_{\tilde \zeta}\} \phi = i ({\cal L}_v-i r v^m a^{(R)}_m)\phi- \frac{i}{2} r s {\cal H} \phi- \frac{i}{2} r \tilde s \widetilde {\cal H} \phi~,\cr
 && \{\delta_\zeta, \delta_{\zeta}\} \phi=0,\qquad \{\delta_{\tilde \zeta}, \delta_{\tilde \zeta}\} \phi=0~.
 \eea
On a round two-sphere of radius $R_{S^2}$ with the standard choice of orthonormal frame:
 \be\label{roundmet}
 ds^2= e^1 e^{\bar 1}~,\qquad e^1= \frac{2 R_{S^2}}{1+z \bar z}dz
 \ee
consider the spinors 
 \bea
 \label{2dspin}
 &&\zeta_*=\begin{pmatrix}
    \zeta_-  \\ \zeta_+
  \end{pmatrix}=\frac{1}{\sqrt{(1+ z \b z)}}\begin{pmatrix}
     \tilde A+i B \b z  \\
  B+i\tilde A z
  \end{pmatrix}~,\nonumber \\
 && \tilde \zeta_*=\begin{pmatrix}
   \tilde \zeta_-  \\ \tilde \zeta_+
  \end{pmatrix}=\frac{1}{\sqrt{(1+ z \b z)}}\begin{pmatrix}
 - \tilde B +i A \b z  \\
    A- i \tilde B z
  \end{pmatrix}~,
 \eea
where $A, B, \tilde A, \tilde B$ are complex constants.

The spinors in~\eqref{2dspin} are solutions of the Killing spinor equations~\eqref{Killing} provided that the background $U(1)_R$ connection vanishes while the two scalars $\cal H$ and $\widetilde{\cal H}$ are constant. 
 \be
 {\cal H}=\frac{i}{R_{S^2}}~,\qquad  \tilde {\cal H}= \frac{i}{R_{S^2}}~.
 \ee  
In the following we will single out two supercharges corresponding to $\zeta$ and $\tilde \zeta$ with $A=\tilde A=1$ and $B=\tilde B=0$.
The spinors bilinears~\eqref{bildef}, built from $\zeta$ and $\tilde \zeta$ are then
\be
 v= {i\over R_{S^2}}(z\partial_z -\bar z \partial_{\bar z })~,\quad s={z \bar z\over 1+ z \bar z}~,\quad \tilde s ={1\over 1+z \bar z}~.
\ee
Near the $z=0$ pole the spinors and the  supercharge corresponding to $\delta_Q=\delta_\zeta+\delta_{\tilde \zeta}$ approach those corresponding to the $\bar A$ topological twist. Similarly near the $z=\infty$ pole the supercharge approaches that corresponding to the $A$ topological twist. More precisely $\delta^2_Q\supset \varepsilon \cal M$, where $\cal M$ is the generator of the U(1) isometry $i(z\partial_z-\bar z \partial_{\bar z})$. Hence near the poles we have an equivariant deformation of the corresponding topological supercharge. By making use of~\eqref{algsimp} and the expression for $v$ we fix the equivariant parameter $\varepsilon$ to be
$$\varepsilon={1\over R_{S^2}}~.$$
We can squash the two sphere maintaining the $U(1)$ isometry generated by $i(z\partial_z-\bar z \partial_{\bar z})$. The maximum number of supercharges is then reduced to two corresponding to two Killing spinors $\zeta$ and $\tilde \zeta$. This squashing is explored in more detail in Appendix~\ref{squash}. 

\subsection{Supersymmetric background \texorpdfstring{$U(1)$}{U1} gauge multiplet}

A two dimensional ${\cal N}=(2,2)$ vector multiplet in Wess-Zumino gauge comprises two scalars $\sigma$ and $\tilde \sigma$, a gauge connection $A_m$,  an auxiliary field $D$ and spinors $\lambda$ and $\tilde \lambda$.
We will consider a background gauge multiplet and set $\lambda=0,~~ \tilde \lambda=0$. 
In the presence of a background gauge multiplet the algebra satisfied by the susy variations on any field $\phi$ of R-charge $r$ and $U(1)$ charge $q$ is modified from~\eqref{algsimp}
\bea
 &&\{\delta_{\tilde \zeta}, \delta_\zeta \}\phi = i ({\cal L}_v-i q v^m A_m-i r v^m a^{(R)}_m)\phi- \frac{i}{2}  s (r {\cal H}+2 q \sigma)\phi- \frac{i}{2}  \tilde s(r \tilde {\cal H}+2 q \tilde \sigma)\phi~,\nonumber \cr
&& \{\delta_\zeta, \delta_{\zeta}\} \phi=0,\qquad \{\delta_{\tilde \zeta}, \delta_{\tilde \zeta}\} \phi=0~.
\eea
On the round sphere we can turn on a background gauge multiplet that preserves all four supercharges~\cite{Closset:2014pda}. This requires that $\sigma$ and $\tilde \sigma$ are constant and:
\be
D=- \frac{1}{2}(\tilde {\cal H} \sigma +{\cal H}\tilde \sigma)~,\qquad F_{1\bar 1}= \frac{i}{4}(\tilde {\cal H} \sigma -{\cal H}\tilde \sigma)~.
\ee
where $F_{m n}$ is the field strength of $A_m$ and $F_{1\bar 1}$ are the corresponding frame components.
We will select
\be\label{gaubckg} A_m=0~,\quad \sigma=i {u\over R_{S^2}} ~,\quad \tilde \sigma=i  {u\over R_{S^2}} ~,\quad D =u\,R^{-2}_{S^2}\ee~where $u$ is a complex constant. More general choices with nonzero flux threading the $S^2$ may also be interesting.
The anticommutator of $\delta_{\tilde \zeta}$ and $\delta_\zeta$ reduces to
\be
 \{\delta_{\tilde \zeta}, \delta_\zeta \}\phi = i {\cal L}_v\phi+ \frac{s+\tilde s}{2R_{S^2}} \left(r+2 q u\right)\phi~.
\ee
The background gauge multiplet is less constrained if we want to preserve only the two supercharges we singled out setting $A=\tilde A=1$ and $B=\tilde B=0$. Introducing a function $f(z \bar z)$ we can then set
\bea\label{bckgh}
&& \quad A=-{i u\over 2}\left(1- f\right)\left({dz\over z}-{d\bar z \over \bar z} \right)~,\cr
&& \sigma=i {u f\over R_{S^2}} ~,\quad \tilde \sigma=i  {u f\over R_{S^2}}~,\quad D= {u f\over R_{S^2}^2}-u{(1- z^2 \bar z^2)\over 2  R_{S^2}^2 } f'~,
\eea
where $u$ is a complex constant as above.
The anticommutator of $\delta_{\tilde \zeta}$ and $\delta_\zeta$ reduces to
\be
 \{\delta_{\tilde \zeta}, \delta_\zeta \}\phi = i {\cal L}_v\phi+ \frac{1}{2R_{S^2}} \left(r+2 q u\right)\phi~.
\ee
Hence the function $f(z \bar z)$ does not appear in the superalgebra. By exploiting the freedom in the choice of $f(z \bar z)$ we can set
\be 
A=-{i u\over 2}\left({dz\over z}-{d\bar z\over \bar z}\right)
\ee
almost everywhere except at the poles of the sphere where $f=1$ to ensure smoothness. In this limit the scalar field $D$ is zero except for two dimensional delta functions supported at the poles of the sphere
\be{}
\label{degdd}
D=2\pi \delta_{np} +  2\pi \delta_{sp}~.
\ee

\subsection{Chiral and antichiral multiplets}
\label{chirachir}
We consider a chiral multiplet of $R$-charge $r$ and vanishing central charges. The multiplet is composed of a scalar $X$ of $R$-charge $r$ a spinor $\psi$ of $R$-charge $r-1$ and an auxiliary scalar field $F$ of $R$-charge $r-2$.  For later use we give charge $q$ to the chiral multiplet under a background $U(1)$ gauge multiplet. For $r=0$, which is the case of our interest, the susy transformations of the components are~\cite{Closset:2014pda}:
\begin{eqnarray}
&&\delta X = 
\sqrt{2}\zeta\psi~, \nonumber \\
&&\delta \psi= \sqrt 2 \zeta F -i \sqrt 2 \gamma^m \tilde \zeta D_{m} X +\frac{i}{\sqrt 2} q (\sigma+\tilde \sigma)\tilde \zeta X+ \frac{i}{\sqrt 2} q (\sigma-\tilde \sigma)\gamma_3 \tilde \zeta X~~,\\
&&\delta F = -i\sqrt{2} D_m (\tilde \zeta\gamma^m \psi) + \frac{i}{\sqrt{2}}({\cal H}+\tilde{\cal H}-q (\sigma+\tilde \sigma))\tilde \zeta\psi- \frac{i}{\sqrt{2}}({\cal H}-\tilde {\cal H}-q (\sigma-\tilde \sigma))\tilde \zeta\gamma_3 \psi~.\nonumber
\end{eqnarray}
The covariant derivatives in the expression above include the $U(1)$ connection $A_m$ and the $U(1)_R$ connection $a^{(R)}_m$.

Similarly an anti-chiral multiplet comprises of a scalar $\tilde X$ of $R$-charge $-r$ a spinor $\tilde \psi$ of $R$-charge $1-r$ and an auxiliary scalar field $\tilde F$ of $R$-charge $2-r$.  For later use we give charge $-q$ to the antichiral multiplet under a background $U(1)$ gauge multiplet. The susy transformations of the components for $r=0$ are~\cite{Closset:2014pda}:
\begin{eqnarray}
&&\delta \widetilde X = \sqrt 2 \tilde \zeta \tilde F +i \sqrt 2 \gamma^m  \zeta D_{m} \widetilde X -\frac{i}{\sqrt 2} q (\sigma+\tilde \sigma) \zeta \widetilde X+ \frac{i}{\sqrt 2} q (\sigma-\tilde \sigma)\gamma_3 \zeta \widetilde X~~,\\
&&\delta \tilde F = -i\sqrt{2} D_m ( \zeta\gamma^m \tilde \psi) + \frac{i}{\sqrt{2}}({\cal H}+\tilde{\cal H}-q (\sigma+\tilde \sigma)) \zeta\tilde \psi+ \frac{i}{\sqrt{2}}({\cal H}-\tilde {\cal H}-q (\sigma-\tilde \sigma)) \zeta\gamma_3 \tilde \psi~.\nonumber
\end{eqnarray}

\subsubsection{Chiral multiplet with gauged isometries}

The scalar fields at the bottom of chiral and anti-chiral multiplets parametrize a sigma model with K\"ahler target space.
If the target space admits a holomorphic Killing vector $k$ 
we can gauge the corresponding isometry by coupling to a vector multiplet. The gauging considered in the previous section corresponds to the Killing vector $k^i=i q_i X^i$. The gauge transformation parameters can be promoted to a chiral superfield $\mathbf \Lambda$. The gauge transformation of a chiral superfield $\mathbf \Phi^i$ is then given by~\eqref{superfields-gauge}:
$$
\delta \mathbf\Phi^i=i \mathbf \Lambda k^i(\mathbf\Phi)~,
$$
Letting $\epsilon$ be the lowest component of $i \mathbf \Lambda$ this results in the following:
\begin{align}
    \delta_\epsilon X^i=&\epsilon\, k^i~,\cr
     \delta_\epsilon \psi^i=&\epsilon \,\partial_j k^i \psi^j~,\cr
     \delta_\epsilon F^i=&\epsilon \,\partial_j k^i F^j+ \frac{1}{2}\epsilon\, 
     \partial_j\partial_k k^i\,\psi^j\psi^k~.
\end{align}
 The auxiliary fields $F^i$ do not transform homogeneously under gauge transformations. We can however define new fields 
\be{}\label{hatF}
\widehat F^i= F^i-{1\over2}\Gamma^i_{j k}\psi^j \psi^k~,
\ee
which transform homogeneously
\be{}
\delta_\epsilon \widehat F^i=\epsilon {\partial_j k^i}\widehat F^j~.
\ee
Note that $\hat F$ is a function both of $X^i$ and $\bar X^{\bar i}$ through $\Gamma^i_{j k}~.$ A second property of the $\hat F^i$ is that, like the $\psi^i$, they transform tensorially under holomorphic coordinate changes $X^i\rightarrow f^i(X)$.

Specializing to a supersymmetric background gauge field and letting the R-charges of the chiral multiplets be $0$ supersymmetry acts as follows
\bea
&&\delta X^i = \sqrt{2}\zeta\psi^i~,\nonumber \\
&&\delta \psi^i= \sqrt 2 \zeta F^i -i \sqrt 2 \gamma^m \tilde \zeta (\partial_{m}X^i -A_m k^i)+\frac{1}{\sqrt 2} (\sigma+\tilde \sigma)\tilde \zeta k^i+\frac{1}{\sqrt 2}  (\sigma-\tilde \sigma)\gamma_3 \tilde \zeta k^i~~,\\
&&\delta F^i = -i\sqrt{2} (\delta^i_j \nabla_m - A_m \partial_j k^i )(\tilde \zeta\gamma^m \psi^j)\cr  &&\qquad+
\frac{i}{\sqrt{2}}(({\cal H}+\tilde{\cal H})\delta^i_j+i (\sigma+\tilde \sigma)\partial_j k^i )\tilde \zeta\psi^j-
\frac{i}{\sqrt{2}}(({\cal H}-\tilde {\cal H})\delta^i_j+i(\sigma-\tilde \sigma) \partial_j k^i)\tilde \zeta\gamma_3 \psi^j~.\nonumber
\eea

\subsection{Lagrangian}

Here we write the Lagrangian of a supersymmetric nonlinear sigma model with K\"ahler target space coupled to background supergravity. We consider the K\"ahler potential to depend on the bottom components $X^i,~\widetilde X^{\bar i}$ of chiral and antichiral superfields of $R$-charge $0$. When the target space admits a $U(1)$ isometry generated by a holomorphic Killing vector field $k$ we couple the model to a background gauge multiplet.

The Lagrangian, written in terms of the K\"ahler metric $g_{i\bar j}$, the curvature $R_{i \bar j k \bar l}$ of the target space and the Hamiltonian $\cal D$ defined in~\eqref{defDD}, is
\bea\label{dchiral}
{\cal L}_{K(\tilde\Phi,\Phi)}~~ =&& ~2 g_{i\bar j}\,\Big( D_m X^i D^m \bar X^{\bar j} +\sigma \tilde \sigma k^i \tilde k^{\bar j}-{\widehat F}^i {\widehat {\widetilde F}} {}^{\bar j}-i  \tilde\psi^{\bar j}\gamma^m {\bf D}_m \psi^i\Big) -2 D {\cal D}\cr
&&-{1\over 2} \big((\sigma+\tilde \sigma) \tilde\psi^{\bar j}\psi^i- (\sigma-\tilde \sigma)\tilde\psi^{\bar j}\gamma_3 \psi^i \big) \left(  \nabla_i k_{\bar j} -  \nabla_{\bar j}k_i\right)\cr
&&- \frac{1}{2}R_{i \bar j k \bar l}(\psi^i\psi^k)(\tilde\psi^{\bar j}\tilde\psi^{\bar l})~,
\eea
where we defined
\begin{align}
& D_m X^i=\partial_{m}X^i -A_m k^i~,\cr
&\nabla_i k^j=\partial_i k^j+\Gamma^j_{i k}k^k~,\cr
&{\bf D}_m \psi^i= (\nabla_m+i a_m^{R})\psi^i-A_m \partial_j k^i \psi^j+\Gamma^i_{j k}\psi^j D_m X^k\cr
& \widehat F^i= F^i-{1\over2}\Gamma^i_{j k}\psi^j \psi^k~,\qquad {\widehat{\widetilde F}}{}^{\bar i}= \widetilde F^{\bar i}-{1\over2}\Gamma^{\bar i}_{\bar j\bar k}\psi^{\bar j} \psi^{\bar k}~.
\end{align}
When the coupling to the background gauge multiplet is turned off, the Lagrangian is a special case of that written in~\cite{Closset:2014pda} for general $R$-charge and central charge assignments of the chiral fields. 
On a round $S^2$ and when the background gauge multiplet is maximally supersymmetric~\eqref{dchiral} coincides with the Lagrangian derived in~\cite{Jia:2013foa}. When the space-time is flat the Lagrangian can be compared to that in~\cite{Baptista:2007ap}.
  
\section{Cohomological description}\label{s:cohomological}
 
In this section we introduce the cohomological description of the ${\cal N}=(2,2)$ supersymmetric non-linear sigma models presented in the previous section. We first review the description of the A-model and then introduce its different equivariant modifications. 
 
\subsection{A-model}\label{ss:a-model}
 
To start we define the A-model (topological sigma model). We follow closely the original works \cite{Witten:1988xj, Witten:1991zz}. We consider only the case when the target manifold $M$ is K\"ahler. 
 
The A-twist of a ${\cal N}=(2,2)$ sigma model can be formulated in terms of the following set of fields: the map $X:\Sigma\rightarrow M$ and $\Psi^\mu$ which is an odd zero form with values in $X^*(TM)$. We also introduce an odd one form $\chi^\mu$ valued in $X^*(TM)$. These one forms $\chi^\mu$ are constrained as we describe below to reduce the number of their components. Let $J$ be a complex structure on $M$. The Hodge star operation $\star$ on $\Sigma$ requires only a complex structure on $\Sigma$ and has the property $\star^2=-1$ when acting on $\Omega^1(\Sigma)$. Thus on one forms with values in $X^*(TM)$ (here we assume  the complexified $TM$) the operation $\star J$ has the property $(\star J)^2=1$~.  Assuming the conventions $\star dz = -i dz$ and $\star d\bar{z} = i d\bar{z}$ the action of $\star J$ on $\Omega^1(\Sigma, X^*(TM))$ is
   \be
    (\star J) \chi^i_z dz = \chi^i_z dz~,~~~
    (\star J) \chi^{\bar{i}}_z dz = -\chi^{\bar{i}}_z dz~,~~~
    (\star J) \chi^i_{\bar{z}} d\bar{z} = - \chi^i_{\bar{z}} d\bar{z}~,~~~
    (\star J) \chi^{\bar{i}}_{\bar{z}} d\bar{z} = \chi^{\bar{i}}_{\bar{z}} d\bar{z}~. 
   \ee
We can introduce two projectors 
  \be
   \frac{1}{2}( 1 \pm \star J)~,
  \ee
which define the notion of (anti) self-duality for $\Omega^1(\Sigma, X^*(TM))$. Let us keep only the fields in the kernel of $1/2(1+\star J)$: $\chi^i_{\bar{z}} d\bar{z} \in \Omega^{0,1}(\Sigma, X^*(T^{1,0}M))$ and  $\chi^{\bar{i}}_{z}dz \in \Omega^{1,0}(\Sigma, X^*(T^{0,1}M))$.  We define the cohomological theory by the following odd transformations which can be interpreted as the de Rham differential on the space of fields
    \be
    \delta X^\mu = \Psi^\mu~,~~~\delta \Psi^\mu = 0~,~~~ \delta \chi^i_{\bar{z}} = h^i_{\bar{z}}~,~~~
    \delta h^i_{\bar{z}}=0 ~,~~~\delta \chi^{\bar{i}}_{z} =  h^{\bar{i}}_{z}~,~~~\delta h^{\bar{i}}_{z}=0 \label{A-model-transf-origianal}~.
    \ee
Here we introduced a one form $h^\mu$ in the kernel of $1/2(1+\star J)$~. The field $h^\mu$ transforms non-tensorially under target space diffeomorphism. Namely if we perform the change of coordinates
   \be
   \tilde{\chi}^i_{\bar{z}} = \frac{\partial \tilde{x}^i}{\partial x^j} \chi^j_{\bar{z}}
   \ee
we have 
   \be
   \tilde{h}^i_{\bar{z}} = \delta \Big (\frac{\partial \tilde{x}^i}{\partial x^j} \chi^j_{\bar{z}} \Big ) = \frac{\partial \tilde{x}^i}{\partial x^j} h^j_{\bar{z}} + \frac{\partial^2 \tilde{x}^i}{\partial x^j \partial x^k} \Psi^j \chi^k_{\bar{z}}~.
   \ee
The field $H_{\bar{z}}^i$ is related to $h_{\bar z}^i$ as follows
   \be
    h^i_{\bar{z}} = H^i_{\bar{z}} -  \Gamma^i_{~jk} \Psi^j \chi^k_{\bar{z}}~,\label{definition-H}
    \ee
where we used the Levi-Civita connection $\Gamma$ for the K\"ahler metric $g$ (similarly for $H^{\bar{i}}_z$). The field $H^\mu$ is also in the kernel of $1/2(1+\star J)$ and transforms tensorially under a target space diffeomorphism, $H^i_{\bar{z}} d\bar{z} \in \Omega^{0,1}(\Sigma, X^*(T^{1,0}M))$ and  $H^{\bar{i}}_{z}dz \in \Omega^{1,0}(\Sigma, X^*(T^{0,1}M))$. 
     
Finally combining (\ref{A-model-transf-origianal}) with the definition (\ref{definition-H}) we obtain the standard formulas for the A-twist of the sigma model
\begin{equation}\label{full-A-model}
\begin{split}
  &  \delta X^\mu=\Psi^\mu \\
  &  \delta \Psi^\nu= 0 \\
  &  \delta \chi^i_{\bar{z}}= H^i_{\bar{z}} - \Gamma^i_{~jk} \Psi^j \chi^k_{\bar{z}}~,\qquad\qquad\qquad\; \delta \chi^{\bar{i}}_{z}= H^{\bar{i}}_{z} - \Gamma^{\bar{i}}_{~\bar{j}\bar{k}} \Psi^{\bar{j}} \chi^{\bar{k}}_{z} \\
  &  \delta H^i_{\bar{z}} =  - \Gamma^{i}_{~jk} \Psi^{j} H^k_{\bar{z}}  + R^i_{~j k\bar{l}} \Psi^k
  \Psi^{\bar{l}} \chi^j_{\bar{z}}~,~~~ \delta H^{\bar{i}}_z =  - \Gamma^{\bar{i}}_{~\bar{j}\bar{k}} \Psi^{\bar{j}}
  H^{\bar{k}}_z  +  R^{\bar{i}}_{~\bar{j} k \bar{l}} \Psi^k \Psi^{\bar{l}} \chi^{\bar{j}}_z~
\end{split}
\end{equation}   
The action of the A-model is defined as the following BRST exact term
   \bea
    S_A = 8 \int \delta \Big ( \chi^i_{\bar{z}}(H_z^{\bar{j}} - \partial_z X^{\bar{j}}) g_{i\bar{j}}
     + \chi^{\bar{j}}_z (H^i_{\bar{z}} - \partial_{\bar{z}}  X^{\bar{j}} ) g_{i\bar{j}} \Big )  dz \wedge d\bar{z} ~.
   \eea
If we integrate out the $H$-field we arrive at the action
     \bea
    S_A =  \int \Big ( g_{\mu\nu}~dX^\mu \wedge \star dX^\nu  + \omega_{\mu\nu} ~dX^\mu \wedge dX^\nu + ...\Big )~, 
     \eea
where dots stand for fermionic terms. The model localizes on the holomorphic maps $\partial_{\bar{z}} X^i=0$. The interesting observables  are labelled by the de Rham cohomology $H_{dR}(M)$. If, as an example, we take a closed two form $\omega$ then we can define the following objects
     \bea\label{obs-amodel}
     && O_0 = \frac{1}{2} \omega_{\mu\nu} (X) \Psi^\mu \Psi^\nu~,\nn\\
     && O_1 = \omega_{\mu\nu} (X) dX^\mu \Psi^\nu~,\\
     && O_2 = - \frac{1}{2} \omega_{\mu\nu}(X) dX^\mu \wedge dX^\nu~,\nn
     \eea
which satisfy the following relations
    \bea
     && \delta O_0 =0~,\nn\\
     && \delta O_1= d O_0~,\\
     && \delta O_2 = d O_1~.\nn
    \eea
This allows us to define the following BRST invariant observables
     \be\label{obs-A-model-extra}
      O_0~,~~~~\int\limits_{\gamma} O_1~,~~~~\int\limits_{\Sigma} O_2~. 
      \ee
Moreover if in the definition (\ref{obs-amodel})  we shift the two form $\omega$ by an exact form  the observables (\ref{obs-A-model-extra}) get shifted by $\delta$-exact terms and thus the deformation will vanish under the path integral. These observables can be generalized to any closed form and up to $\delta$-exact terms they depend only on the corresponding class in $H_{dR}(M)$ (see \cite{Witten:1988xj, Baulieu:1989rs, Witten:1991zz} for further explanation). The correlators in the A-model are related to the deformation of the ring structure of $H_{dR}(M)$ and this cohomological model is the physical counterpart of the Gromov-Witten theory.  
    
In what follows we use the following terminology: the A-model is defined as above with the $\chi$ and $H$ fields in the kernel of $\frac{1}{2}(1+\star J)$. The model is localized on holomorphic maps $\partial_z X^{\bar{i}}=0$.  We refer to  the $\bar{\rm A}$-model with the $\chi$ and $H$ fields in the kernel of $\frac{1}{2}(1-\star J)$. The model is then localized on anti-holomorphic maps $\partial_z X^{i}=0$.

\subsection{New self-duality on \texorpdfstring{$S^2$}{S2}}\label{ss:projector}

As we reviewed in the previous subsection, the construction of the A-model uses the notion of (anti)self-duality on $\Omega^1(\Sigma, X^*(TM))$ which is based on the operator $\star J$ satisfying $(\star J)^2=1$. In 4D it has been pointed out in \cite{Festuccia:2018rew} that if a manifold is equipped with a vector field $v$ then one can generalize the notion of (anti)self-duality for two forms. Formally we can repeat this construction for the 2D case on one-forms. Assume that $\Sigma$ has a vector field $v$ and a metric $g$ such that $\kappa=g(v)$ (with $||v||^2 = \iota_v \kappa$).  Away from the fixed points of $v$ we can define the following operation on $\Omega^1(\Sigma, X^*(TM))$ 
    \bea
     m = \Big (-1 + 2 \frac{\kappa \wedge \iota_v}{||v||^2} \Big )~, 
    \eea
which acts on target space indices as the identity. This operation has the following property
  \bea
   m^2 = 1~,~~~~\star m + m \star =0~. 
  \eea
Thus away from the fixed point of $v$ we can define a new operation $\alpha \star J + \beta m$ such that
   \bea
    (\alpha \star J + \beta m)^2=1~,
   \eea 
with $\alpha^2+\beta^2=1$. Choosing $\alpha = \cos 2\rho$ and $\beta = \sin 2\rho$ with $\rho$ being a suitably chosen function on $\Sigma$ (see section 2.1 in \cite{Festuccia:2018rew} for other formal considerations which are equally applicable here) we can construct a new projector 
   \bea
    P^+ = \frac{1}{2} ( 1 + \cos 2\rho \star J + \sin 2\rho~ m)~,
   \eea
which can be smoothly extended to the fixed points $v$. There are further generalizations where in front of $m$ instead of the identity tensor we put another tensor on $M$ squaring to 1.
 
Consider specifically the case of $S^2$ (see Appendix \ref{app:S2} for our conventions on $S^2$). On $S^2$ we choose the vector field $v$ associated with a $U(1)$ rotation of $S^2$. In what follows we use two sets of coordinates: the spherical coordinates $(\theta, \phi)$ and the stereographic coordinates $(z,\bar{z})$ (or $(z',\bar{z}')$ in another patch).  If we choose the function $\rho$ such that
  \be
  1- \sin 2\rho = \frac{2}{1+ \cos^2 \theta}
  \ee
then we get the following projector\footnote{There exists another set of projectors based on $\frac{1}{1+\cos^2 \theta} (1 + \cos \theta~ \star J - \kappa\wedge \iota_v)$.} 
  \be
  P^+ = \frac{1}{1+\cos^2 \theta} (1 - \cos \theta~ \star J - \kappa\wedge \iota_v)=
  \frac{1}{1+\cos^2 \theta} (1 - \cos \theta~ \star J - \sin^2 \theta~ d\phi \wedge \iota_{\partial_\phi})~,\label{P+spheric-first}
  \ee
where we have used $||v||^2 = \sin^2 \theta$. This concrete form of the projector is motivated by supersymmetry considerations. The projector $P^+$ simplifies to $\frac{1}{2}(1+\star J)$ at the north pole $\theta=\pi$ and to to $\frac{1}{2}(1-\star J)$ at the south pole $\theta =0$~. We also define the complementary projector $P^-=1-P^+$ and decompose one forms with values in $X^*(TM)$ as follows
  \be\label{def-omega1TM}
   \Omega^1(S^2, X^*(TM))= \Omega^{1+}(S^2, X^*(TM)) \oplus \Omega^{1-}(S^2, X^*(TM))
  \ee
where
   $$\Omega^{1\pm} (S^2, X^*(TM))= P^{\pm} \Big (\Omega^1(S^2, X^*(TM)) \Big )~.$$
Some explicit formulas for this decomposition are as follows. In $(z,\bar{z})$ coordinates the projector (\ref{P+spheric-first}) is written as
 \be
 \label{projzzb}
   P^+ = \frac{1}{2(1+|z|^4)} \Big ( 1 + |z|^4 + (1-|z|^4) \star J 
    + 2 (z^2 d\bar{z}\wedge \iota_{\partial_z} + \bar{z}^2 dz \wedge \iota_{\partial_{\bar{z}}}) \Big )~. 
 \ee
Consider  
$$\chi^\mu = \chi^\mu_z dz + \chi^\mu_{\bar{z}} d\bar{z} \in \Omega^1(S^2, X^*(TM))~.$$
If $\chi \in \Omega^{1-}(S^2, X^*(TM))$ then 
 \be
  P^+ \chi=0~~~\rightarrow~~~\chi^i_z + \bar{z}^2 \chi^i_{\bar{z}}=0~,~~\chi^{\bar{i}}_{\bar{z}} + z^2 \chi^{\bar{i}}_{z}=0
 \ee
and the basis for $\Omega^{1-}(S^2, X^*(TM))$ is given by
  \be
   \chi^i_{\bar{z}}(d\bar{z} - \bar{z}^2 d z)~,~~~~ \chi^{\bar{i}}_{z}(dz - z^2 d \bar{z})~.
  \ee
Let us consider the differential equation
 \be
  P^+ dX^i=0~~~\rightarrow~~~X^i\Big (\frac{\bar{z}}{1+|z|^2}\Big )
 \ee
where $X^i$ is an holomorphic map from the disk to $M$. We will elaborate on this point later on. 
 
On the other hand for $\chi \in \Omega^{1+}(S^2, X^*(TM))$ 
    \be
  P^- \chi=0~~~\rightarrow~~~\chi_{\bar{z}}^i - z^2 \chi^i_{z}=0~,~~
  \chi^{\bar{i}}_{z} - \bar{z}^2 \chi^{\bar{i}}_{\bar{z}}=0
 \ee
and the basis for $\Omega^{1+}(S^2, X^*(TM))$ is given by 
\be
 \chi^i_z (dz + z^2 d\bar{z})~,~~~~\chi^{\bar{i}}_{\bar{z}} (d\bar{z} + \bar{z}^2 dz)~. 
\ee
The PDE with $P^-$ projector can be reduced to
 \be
  P^- dX^i=0~~~\rightarrow~~~X^i\Big (\frac{z}{1-|z|^2}\Big )
 \ee
where $X^i$ is a holomorphic map from the disk to $M$.

 \subsection{New model on $S^2$}
 \label{ss:new-model-1}
 
In this subsection we present a new cohomological field theory on $S^2$ which is a reformulation of the supersymmetric theory in terms of differential forms. There are two novel aspects of our construction: we introduce equivariance with respect to the rotation of $S^2$ and we use the exotic projector $P^+$ defined in the previous subsection.
   
As for the A-model we define the space of maps $X:S^2\rightarrow M$, an odd zero form $\Psi^\mu$ with values in $X^*(TM)$ and  $\chi^\mu$ an odd element of $\Omega^{1+}(S^2, X^*(TM))$. The cohomological transformations are defined as 
\begin{equation}
\label{cohomj}
\begin{split}
  &  \delta X^\mu=\Psi^\mu \\
  &  \delta \Psi^\nu= \mathcal{L}_v X^\nu \\
  &  \delta \chi^\mu= h^\mu \\
  &  \delta h^\mu = {\cal L}_v \chi^\mu \\
\end{split}
\end{equation}
where we introduced $h^\mu$ as partner for $\chi^\mu$. The transformations are such that $\delta^2={\cal L}_v$. 
 In front of $v$ we assume the equivariant parameter\footnote{Please note that we use $\varepsilon$ for 
  the equivariant paramater on $S^2$ and $\e^a$ as equivariant parameters for a target manifold.} $\varepsilon$. However in most of 
  the formulas we will not write $\varepsilon$ to avoid unreadable expressions unless it is crucial for the discussion like in section \ref{s:1-loop}. 
  As before the new field  $h^\mu$ does not transform tensorially under target space diffeomorphism and to fix it we introduce 
 \be
  h^\mu = H^\mu - \Gamma^\mu_{~\nu\rho} \Psi^\rho \chi^\nu~,
  \ee
where $H^\mu$ is an even element of  $\Omega^{1+}(S^2, X^*(TM))$. Next we calculate 
    $$\delta H^\mu = {\cal L}_v \chi^\mu  + \delta ( \Gamma^\mu_{~\nu\rho} \Psi^\rho \chi^\nu) = {\cal L}_v \chi^\mu  + \Gamma^\mu_{~\nu\rho} {\cal L}_v X^\rho \chi^\nu
    - \Gamma^{\mu}_{~\nu\rho} \Psi^{\nu} H^\rho  + \frac{1}{2} R^\mu_{~\nu\rho\sigma} \Psi^\rho\Psi^\sigma \chi^\nu $$
We define the covariant version of the Lie derivative 
    $$ {\cal L}^{\Gamma}_v \chi^\mu=  {\cal L}_v \chi^\mu +  \Gamma^\mu_{~\nu\rho} {\cal L}_v X^\rho \chi^\nu~,$$
which transforms tensorially under target space diffeomorphism and rewrite the transformations as follows
  \begin{equation}\label{transf-model-1}
\begin{split}
  &  \delta X^\mu=\Psi^\mu \\
  &  \delta \Psi^\nu= \mathcal{L}_v X^\nu \\
  &  \delta \chi^\mu= H^\mu - \Gamma^\mu_{~\nu\rho} \Psi^\rho \chi^\nu  \\
  &  \delta H^\mu = {\cal L}^\Gamma_v \chi^\mu  - \Gamma^{\mu}_{~\nu\rho} \Psi^{\nu} H^\rho  + \frac{1}{2} R^\mu_{~\nu\rho\sigma} \Psi^\rho\Psi^\sigma \chi^\nu 
\end{split}
\end{equation}   
Here we prescribe that the fields $\chi$ and $H$ are in the image of $P^+$. This cohomological description is a reformulation of ${\cal N}=(2,2)$ theory on $S^2$ described in Section \ref{s:sigma-S2}. An explicit map between ${\cal N}=(2,2)$ fields and the cohomological variables~\eqref{cohomj} is presented in~\eqref{cohdef}~.
    
An important feature of this model derives from the following property of the BRST-exact term    
    \be\label{S2-exact-explicit}
    S= \int \delta \Big ( 4 (1+ \cos^2 \theta) \chi^\mu \wedge \star (dX^\nu - H^\nu) g_{\mu\nu} +
     \Psi^\mu \wedge \star {\cal L}_v X^\nu g_{\mu\nu} \Big ) 
    \ee
which can be rewritten as follows
\begin{equation}
\begin{split}\label{new-eaxct-action}
 &    S= \int \Big [ (1+\cos^2 \theta) (P^+ dX)^\mu \wedge \star (P^+ dX)^\nu g_{\mu\nu} + {\cal L}_v X^\mu \wedge \star {\cal L}_v X^\mu g_{\mu\nu}  + ... \Big ] \\
 &  = \int \Big [ dX^\mu \wedge \star dX^\nu g_{\mu\nu} + \cos \theta~ \omega_{\mu\nu} dX^\mu \wedge dX^\nu + ... \Big ]~.
\end{split}
\end{equation} 
Here we used $\star 1= \Omega_2$ and the dots stand for fermionic terms. Thus the supersymmetric sigma model action is BRST-exact up to the term $\cos \theta ~X^*(\omega)$ which can be thought of as an equivariant analog of the topological term $X^*(\omega)$. We will discuss this term and its generalizations when we consider observables in section \ref{s:observables}. To clarify the meaning of our theory we can rewrite the bosonic integrand in (\ref{new-eaxct-action}) in mixed coordinates (using both $(z,\bar{z})$ and $(\theta, \phi)$) as follows
\begin{equation}
\begin{split}
  &  dX^\mu \wedge \star dX^\nu g_{\mu\nu} + \cos \theta~ \omega_{\mu\nu} dX^\mu \wedge dX^\nu\\
 &  = 4i \Big ( \cos^2 \frac{\theta}{2}~\partial_{\bar{z}} X^i \partial_z X^{\bar{j}} 
    + \sin^2 \frac{\theta}{2}~ \partial_z X^i \partial_{\bar{z}} X^{\bar{j}} \Big )g_{i\bar{j}} dz \wedge d\bar{z}~.
  \end{split}
\end{equation} 
This form is very reminiscent of the formulas in the 4D theory of Pestun on $S^4$ \cite{Pestun:2007rz}. We see that holomorphic maps dominate around $\theta=0$ while anti-holomorphic maps dominate around~$\theta=\pi$.

 \subsection{New model with target-space equivariance}\label{ss:model-2}
    
If the target space $M$ admits a torus action then there exists a natural extension of the model introduced in the previous section. Let the holomorphic vector fields $k^\mu_a$ generate a torus action on K\"ahler manifold $M$ and preserve the K\"ahler structure, i.e. ${\cal L}_{k_a} g=0$, ${\cal L}_{k_a} \omega=0$ and ${\cal L}_{k_a} J=0$. Moreover we require that this action is Hamiltonian 
  \be
   k^\mu_a \omega_{\mu\nu} + \partial_\nu {\cal D}_a=0~. \label{eq-Hamiltonian} 
  \ee
As before  we define the space of maps $X:S^2\rightarrow M$, an odd zero form $\Psi^\mu$ with values in $X^*(TM)$ and  $\chi^\mu$ an odd element of $\Omega^{1+}(S^2, X^*(TM))$. The cohomological transformations are now defined as follows 
 \begin{equation}
 \label{cohomequiv}
\begin{split}
  &  \delta X^\mu=\Psi^\mu \\
  &  \delta \Psi^\nu= \mathcal{L}_v X^\nu + \epsilon^a k^\mu_a\\
  &  \delta \chi^\mu= h^\mu \\
  &  \delta h^\mu = {\cal L}_v \chi^\mu  +\epsilon^a \partial_\nu k^\mu \chi^\nu \\
\end{split}
\end{equation}
where we introduce the equivariant parameters $\e^a$ which we assume to be real for the moment. By construction $\delta^2$ is given by a rotation of $S^2$ plus a rotation generated by $k_a$ on the target space $M$.  Using the same logic as before we define the new field 
     \be
      h^\mu = H^\mu - \Gamma^\mu_{~\nu\rho} \Psi^\rho \chi^\nu~,
      \ee
so that $H^\mu$   is even element of $\Omega^{1+}(S^2, X^*(TM))$.  Calculating its transformation results in 
    \bea
   && \delta H^\mu = {\cal L}_v \chi^\mu +  \epsilon^a \partial_\nu k^\mu_a \chi^\nu + \delta ( \Gamma^\mu_{~\nu\rho} \Psi^\rho \chi^\nu) \nn \\
  &&    = {\cal L}_v \chi^\mu  +  \epsilon^a \partial_\nu k^\mu \chi^\nu  + 
   \Gamma^\mu_{~\nu\rho} {\cal L}_v X^\rho \chi^\nu - \epsilon^a \Gamma^\mu_{~\nu\rho} k_a^\rho \chi^\nu 
    -\Gamma^{\mu}_{~\nu\rho} \Psi^{\nu} H^\rho  + \frac{1}{2} R^\mu_{~\nu\rho\sigma} \Psi^\rho\Psi^\sigma \chi^\nu \eea
 so that we obtain the final result
       \begin{equation}\label{def-trans-model2}
\begin{split}
  &  \delta X^\mu=\Psi^\mu \\
  &  \delta \Psi^\nu= \mathcal{L}_v X^\nu + \epsilon^a k^\mu_a\ \\
  &  \delta \chi^\mu= H^\mu - \Gamma^\mu_{~\nu\rho} \Psi^\rho \chi^\nu  \\
  &  \delta H^\mu = {\cal L}^\Gamma_v \chi^\mu  + \epsilon^a (\nabla_\nu k_a^\mu )\chi^\nu-
   \Gamma^{\mu}_{~\nu\rho} \Psi^{\nu} H^\rho  + \frac{1}{2} R^\mu_{~\nu\rho\sigma} \Psi^\rho\Psi^\sigma \chi^\nu~, 
\end{split}
\end{equation}   
with  
$$ \nabla_\nu k_a^\mu= \partial_\nu k^\mu_a + \Gamma^\mu_{~\nu\rho} k_a^\rho   \ .$$~
As before the fields $\chi$ and $H$ are in the image of $P^+$. In Appendix~\ref{seccoh} we show that this cohomological description is a reformulation of the ${\cal N}=(2,2)$ theory coupled to a background gauge field on $S^2$ described in Section \ref{s:sigma-S2}. 

We can study the BRST-exact term     \be\label{model2-exact-action}
    S= \int \delta \Big ( 4 (1+ \cos^2 \theta) \chi^\mu \wedge \star (dX^\nu + \e^a k_a^\nu A - H^\nu) g_{\mu\nu} +
     \Psi^\mu \wedge \star ({\cal L}_v X^\nu + \e^a k_a^\nu) g_{\mu\nu} \Big ) 
    \ee
where we have introduced a flat background connection
 \be\label{def-connection}
     A= - \frac{i dz}{2 z} + \frac{i d\bar{z}}{2\bar{z}}
    \ee
with singularities at the poles of $S^2$ and the property $\iota_v A=1$. 
We concentrate our attention only on the bosonic terms of this BRST-exact action. Upon the integration of $H^\mu$ we obtain the following bosonic action
 \bea
    &   (1+\cos^2 \theta) [P^+ (dX^\mu + \e^a k^\mu_a A) \wedge \star P^+ (dX^\nu + \e^b k_b^\nu A)  g_{\mu\nu}] \nn \\
    & + ({\cal L}_v X^\mu + \e^a k_a^\mu) \wedge \star 
   ({\cal L}_v X^\mu + \e^b k_b^\nu)  g_{\mu\nu} =   \\
 &  = (dX^\mu + \e^a k^\mu_a A)  \wedge \star ( dX^\nu + \e^b k_b^\nu A) g_{\mu\nu} + \cos \theta~ \omega_{\mu\nu} (dX^\mu + \e^a k^\mu_a A)  \wedge (dX^\nu + \e^b k_b^\nu A) ~,\nn
  \eea
where we have suppressed the integral sign. The last term can be simplified using (\ref{eq-Hamiltonian})
  \bea
&   \cos \theta~ \omega_{\mu\nu} (dX^\mu + \e^a k^\mu_a A)  \wedge (dX^\nu + \e^b k_b^\nu A) =    \cos \theta~ \omega_{\mu\nu} dX^\mu   \wedge dX^\nu 
    + 2    \cos \theta~ \omega_{\mu\nu} \e^a k^\mu_a A \wedge dX^\nu  \nn  \\
&    =   \cos \theta~ \omega_{\mu\nu} dX^\mu   \wedge dX^\nu  - 2 \cos \theta~\e^a A \wedge d{\cal D}_a~.
\eea
We can rewrite the second as follows
  \be
    2 \cos \theta~\e^a A \wedge d{\cal D}_a =  - d (2 \cos \theta A \e^a {\cal D}_a) - 2 \iota_v \Omega_2 \wedge A \e^a {\cal D}_a +
    2 \cos \theta~ dA \e^a {\cal D}_a~,
  \ee
where $\Omega_2$ is defined in (\ref{def-Omega2}), in Appendix. 
Under the integral the first term on the RHS vanishes due to Stokes theorem (the singularity for $A$ does not spoil Stokes theorem). Remember that $A=d\phi$ and thus we have
   \be
    d A = 2\pi \delta_{np} -  2\pi \delta_{sp}~,
   \ee
where on the RHS we have delta function contributions from the north and south poles.  Combing everything together we obtain 
    \be
   \int  2 \cos \theta~\e^a A \wedge d{\cal D}_a =  \int 2 \Omega_2 \e^a {\cal D}_a + 4\pi \e^a [{\cal D}_a (X_{np}) + 
   {\cal D}_a (X_{sp})]
    \ee
where $X_{n.p}$ and $X_{s.p}$ are the values of a map $X$ at north and south poles. Thus to summarize we get the following relation 
\bea
\label{idwtdel}
   S  = ||dX + \e^a k_a A||^2  - \int \Big ( ( \cos \theta + \Omega_2)( -\omega_{\mu\nu} dX^\mu   \wedge dX^\nu
  + 2 \e^a {\cal D}_a ) \Big ) \nn  - \\ - 4\pi \e^a [{\cal D}_a (X_{np}) + 
   {\cal D}_a (X_{sp})] +  ...
\eea
 where the dots stand for the fermionic part. The second term corresponds to the bosonic part of a cohomological observable which we will discuss in detail in the next section. This identity can be used to rewrite the Lagrangian~\eqref{dchiral} in cohomological variables for the special case where the function $f$ specifying the background gauge field configuration~\eqref{bckgh} is taken to vanish except at the poles. In this case the background field $D$ vanishes except for delta functions at the poles as seen in~\eqref{degdd}. These delta functions then cancel the third term in~\eqref{idwtdel}. The rewriting of the action stemming from~\eqref{dchiral} using cohomological variables is considered for generic $f$ in appendix~\ref{seccoh}.
 
 Let us stress one important point. In A-model considerations in subsection \ref{ss:a-model} we could introduce the equivariance in the transformations \eqref{full-A-model}, however 
  it will lead to the modification in the kinetic term (it will not have a canonical form). While in the model descirbed in this subsection and in the previous subsection 
   we have canonical kinetic term and this is due to the natural compatibility 
    properties between the new projector $P^+$ and the equivariance.

 \section{Observables in new model}\label{s:observables} 
 
In this section we study observables in the supersymmetric sigma model on $S^2$. For this discussion we use the cohomological variables although the results can be restated in terms of the original supersymmetry fields from section \ref{s:sigma-S2}. Our present discussion is a generalization of standard A-model considerations from \cite{Witten:1988xj, Baulieu:1989rs, Witten:1991zz}. In subsection \ref{ss:a-model} we reviewed the A-model transformations and the A-model observables (\ref{obs-amodel})-(\ref{obs-A-model-extra}) which are naturally associated with the de Rham cohomology $H_{dR}(M)$ of the target space manifold $M$. 
 
 \subsection{With equivariance on $S^2$}
 
We start by analyzing the observables in the model presented in subsection \ref{ss:new-model-1} where only the equivariance on $S^2$ is present. Using the transformations (\ref{transf-model-1}) we know that $\delta^2 = {\cal L}_v = d_v^2$ on all fields, where $d_v = d + \iota_v$ is the equivariant differential on $S^2$. To be concrete let us pick up an equivariantly closed form $d_v (\Omega_0 + \Omega_2)=0$ on $S^2$ and a closed two form $\omega$ on $M$. We can then introduce the following objects 
  \bea\label{obs-model-equiv-1}
     && O_0 =  \frac{1}{2} \Omega_0~ \omega_{\mu\nu} (X) \Psi^\mu \Psi^\nu~,\nn\\
     && O_1 = \Omega_0 ~\omega_{\mu\nu} (X) dX^\mu \Psi^\nu~,\\
     && O_2 = - \frac{1}{2} \Omega_0 ~\omega_{\mu\nu}(X) dX^\mu \wedge dX^\nu + \frac{1}{2} \Omega_2~
     \omega_{\mu\nu} \Psi^\mu \Psi^\nu~.\nn
  \eea
Using the transformations (\ref{transf-model-1}) we get 
  \bea
   && \delta O_0 =  \iota_v O_1~, \nn  \\
   &&  \delta O_1 = d O_0 +  \iota_v O_2 ~,\\
   && \delta O_2 =d O_1~, \nn
  \eea
which can be  written more compactly as
  \bea
   \delta (O_0 + O_1 + O_2) = d_v (O_0 + O_1 + O_2)~.
  \eea
Thus we conclude that the zero form $O_0$ is a local observable if it is placed either on the north pole $O_0^{np}$ or the south pole $O_0^{sp}$ of $S^2$ where $v$ vanishes and $\delta O_0^{np}=0$, $\delta O_0^{sp}=0$. 
   
The one dimensional integral of the one form $O_1$
   \be
    \int\limits_\gamma O_1 ~,
   \ee
is an observable (i.e., it is annihilated by $\delta$) if $\gamma$ is  a closed 1-dimensional $U(1)$ orbit on $S^2$ (e.g., the equator of $S^2$). Moreover if we change $\gamma$ within homology class 
\be
  \int\limits_\gamma O_1 -  \int\limits_{\tilde{\gamma}} O_1 = \delta \int\limits_{\Sigma} O_2
\ee
 such that $\partial \Sigma = \gamma - \tilde{\gamma}$. Thus on $S^2$ there are no non-trivial observables 
  associated the above integral over $\gamma$. 
Finally the integral over the two form
   \be\label{non-local-general}
      \int\limits_{S^2} O_2 
   \ee
is an observable in our theory. These observables depend on the choice of equivariantly closed form $\Omega_0+\Omega_2$ on $S^2$ and a closed form $\omega$ on $M$. If we shift the equivariantly closed form $\Omega_0+\Omega_2$ as follows
  \be
   \Omega_0+ \Omega_2~\rightarrow \Omega_0 + \Omega_2 + d_v \Omega_1
  \ee
for some $U(1)$-invariant one form $\Omega_1$ on $S^2$ we obtain the shifts 
   \bea
   && O_0~\rightarrow~ O_0 + \iota_v \Big (\frac{1}{2}\Omega_1~ \omega_{\mu\nu} \Psi^\mu \Psi^\nu \Big )~,\nn\\
   && O_1~\rightarrow~ O_1 + \delta  \Big (\frac{1}{2}\Omega_1~ \omega_{\mu\nu} \Psi^\mu \Psi^\nu \Big ) + \iota_v \Big  (\Omega_1 \omega_{\mu\nu} dX^\mu \Psi^\nu \Big )~,\\
   && O_2~\rightarrow~O_2 + \delta \Big (\frac{1}{2} \Omega_1 \omega_{\mu\nu} dX^\mu \Psi^\nu \Big ) + d \Big ( \frac{1}{2} \Omega_1 \omega_{\mu\nu} \Psi^\mu \Psi^\nu \Big )~,\nn
   \eea
that imply the following shifts for the local observables
    \be
    O_0^{np}~\rightarrow~O_0^{np}~,~~~~~~~O_0^{sp}~\rightarrow~O_0^{sp}~,
    \ee
while for the non-local observables
    \be
    \int\limits_{\gamma} O_1~\rightarrow~ \int\limits_{\gamma}O_1 + \delta  \Big (\int\limits_{\gamma}\frac{1}{2}\Omega_1~ \omega_{\mu\nu} \Psi^\mu \Psi^\nu \Big )~,
    \ee
and 
     \be
     \int\limits_{S^2} O_2~\rightarrow~\int\limits_{S^2}O_2 + \delta \Big (\int\limits_{S^2}\frac{1}{2} \Omega_1 \omega_{\mu\nu} dX^\mu \Psi^\nu \Big )~.
    \ee
Thus these observables depend only on the class of $(\Omega_0 + \Omega_2)$ within the equivariant cohomology~$H_{equiv}(S^2)$. 

If instead we shift $\omega$ by an exact form: $\omega \rightarrow \omega + d \nu$ we get the following shifts
     \bea
       && O_0~\rightarrow~ O_0 + \delta (\Omega_0 \nu_\mu \Psi^\mu) - \iota_v (\Omega_0 \nu_\mu dX^\mu)~,\nn\\
        && O_1~\rightarrow~ O_1 + d(\Omega_0 \nu_\mu \Psi^\mu) + \iota_v (\Omega_2 \nu_\mu \Psi^\mu ) - \delta (\Omega_0 \nu_\mu dX^\mu) ~,\\
        && O_2~\rightarrow~O_2 - d (\Omega_0 \nu_\mu dX^\mu) + \delta (\Omega_2 \nu_\mu \Psi^\mu) ~,\nn
     \eea
Thus under such shift the local  observables are shifted by $\delta$-exact terms
     \be
      O_0^{np}~\rightarrow~O_0^{np} +\delta (\Omega_0\nu_\mu \Psi^\mu)^{np}~,~~~~~
      O_0^{sp}~\rightarrow~O_0^{sp} +\delta (\Omega_0\nu_\mu \Psi^\mu)^{sp}~,
     \ee
where $np/sp$ indicates that the expression is evaluated either on the north pole or the south pole. Non-local observables are shifted as follows
     \be
      \int\limits_{\gamma} O_1~\rightarrow~\int\limits_{\gamma} O_1 - \delta \Big ( \int\limits_{\gamma} \Omega_0 \nu_\mu dX^\mu \Big )
     \ee
and 
     \be
      \int\limits_{S^2} O_2~\rightarrow~\int\limits_{S^2} O_2 + \delta \Big ( \int\limits_{S^2} \Omega_2 \nu_\mu \Psi^\mu  \Big )~.
     \ee
Hence these observables depend only on the class of $\omega$ within $H_{dR}(M)$ cohomology. The discussion above can be generalized to any class in $H_{equiv}(S^2)$ and $H_{dR}(M)$ and analyzed in similar fashion and the result is that the cohomology classes of the observables under the operator $\delta$ depend only on the equivariant cohomology on $S^2$, $H_{equiv}^\bullet(S^2)$ and the de Rham cohomology on $M$, $H^\bullet_{dR} (M)$. If we focus on the degree two part of $H_{equiv}^\bullet(S^2)$, which is  two dimensional for $S^2$, we can say that our set of observables corresponds to two copies of $H_{dR}(M)$, one copy is associated with the north pole and another with the south pole. 
     
As we show in appendix~\ref{seccoh} the partition function on $S^2$ can be understood as the evaluation of the non-local observable (\ref{non-local-general}) with the following data:  a K\"ahler form $\omega$ on $M$ and a concrete choice of the equivariantly closed form of degree 2
  \be\label{2-form-sphere}
   \Omega_0 + \Omega_2 = \cos \theta + \sin \theta ~d \phi \wedge d \theta~,
  \ee
which is an equivariant extension of the Fubini-Study volume form.  Within the cohomology class in $H_{equiv}(S^2)$ we can change the representative as 
   \be\label{class-2form}
    \Omega_0 + \Omega_2 + d_v \Omega_1 = (\cos \theta + f (\theta)) + d (\cos \theta + f(\theta)) \wedge d\phi~,
   \ee
where $\Omega_1 = f(\theta) d\phi$ is an invariant one form. Since $d\phi$ is not well defined at the poles we have to require that $f(0)=0$ and $f(\pi)=0$ in order for $\Omega_1$ to be well-defined. From (\ref{class-2form}) we see that within the cohomology class we can set both zero and two forms to zero everywhere 
 except at the poles of $S^2$. Thus cohomologically we can represent (\ref{2-form-sphere}) as
  \be
    \Omega_0 + \Omega_2 =  \Omega_0^N + \Omega_2^N +  \Omega_0^S + 
    \Omega_2^S 
  \ee
with $\Omega_0^N=1$ on north pole and zero elsewhere, $\Omega_0^S=-1$ and zero elsewhere. The partition function on $S^2$ corresponds to the evaluation of the following observable
   \be
  Z_{S^2} = \Big \langle \exp\left[ \int (\Omega_0 + \Omega_2)(-X^*(\omega) + \omega_{\mu\nu} \Psi^\mu \Psi^\nu ) - i \int X^*(b) \right] \Big \rangle 
  \ee
where under the integral we put $2(O_0 + O_2)$ and  we added a $b$-field contribution where $db=0$. The term with the $b$-field
 \be
 \int X^*(b) 
 \ee
can be interpreted as a cohomological observable fitting the description (\ref{non-local-general}) for the choice of $\Omega_0=1$ (the trivial element in $H_{equiv}(S^2)$). Under the integral we can perform the following substitution
   \be
 - \int X^*(b) =\int (\Omega^N_0 - \Omega_0^S + \Omega^N_2 - \Omega_2^S)(-X^*(b) + b_{\mu\nu} \Psi^\mu \Psi^\nu )
 \ee
since we can change the class in $H_{equiv}(S^2)$ without changing the result of integration. After some more manipulations we can rewrite the partition function as the expectation value of the following observables
   \bea\label{observ-non-eq}
  Z_{S^2} = &&\Big \langle \exp\left[ i \int (\Omega^N_0 + \Omega^N_2)(-X^*(-i\omega+ b) + (-i\omega+b)_{\mu\nu} \Psi^\mu \Psi^\nu ) - \right.\nn \\
  && \left. -  i\int (\Omega^S_0 + \Omega^S_2)(-X^*(i\omega+ b) + (i\omega+b)_{\mu\nu} \Psi^\mu \Psi^\nu )\right] \Big \rangle ~.
  \eea
Hence we see that formally the partition function ''factorizes'' in two contributions, one coming from north hemisphere responsible for the $(b-i\omega)$ dependence and another coming from the south hemisphere responsible for the $(b+ i\omega)$ dependence. So far we did not analyze actual localization locus but it should comprise two disks, weighted differently, one going to the north hemisphere and one going to the south hemisphere. At this point this argument is formal and we cannot say that the model simply factorizes in two copies of the A-model and the $\bar{\rm A}$-model for two disks. The equivariance of $S^2$ plays non-trivial role and we will see this even at the level of the constant map contribution.

 \subsection{Target space equivariance}
 
Here we consider the model with both equivariance on $S^2$ and equivariance on $M$. This model has been described in subsection \ref{ss:model-2} and it is defined by the cohomological transformations (\ref{def-trans-model2}). For the sake of clarity let us focus on specific observables related to the K\"ahler form $\omega$ and its equivariant extension on $M$ given by
   \be
  \iota_{k_a} \omega + d{\cal D}_a =0~,
 \ee
where ${\cal D}_a$ is a Hamiltonian for $k_a$. Picking an equivariant closed form $(\Omega_0 + \Omega_2)$ on $S^2$ we can define 
\bea\label{model2-obs-def}
  && O_0 =  \frac{1}{2} \Omega_0~ \omega_{\mu\nu} \Psi^\mu \Psi^\nu + \Omega_0 \e^a {\cal D}_a ~,\nn\\
  && O_1 =  \Omega_0 ~\omega_{\mu\nu} dX^\mu \Psi^\nu     ~,\\
  && O_2 = - \frac{1}{2}\Omega_0~  \omega_{\mu\nu} dX^\mu \wedge dX^\nu + \Omega_2 \e^a {\cal D}_a  + \frac{1}{2} \Omega_2~
   \omega_{\mu\nu} \Psi^\mu \Psi^\nu~.\nn
\eea
Using the transformations (\ref{def-trans-model2}) we derive the relations
  \bea\label{model2-obs-rel}
 && \delta O_0 =  \iota_v O_1~, \nn  \\
&&  \delta O_1 = d O_0 +  \iota_v O_2 ~,\\
&& \delta O_2 =d O_1~. \nn
 \eea
Following the logic described previously we can define the local observables: $O_0^{np}$ and $O_0^{sp}$ ($O_0$ evaluated at either at the north pole or at the south pole) and non-local observables: the integral of $O_1$ over a closed invariant $U(1)$ orbit and the integral of $O_2$ over $S^2$. If in the definition (\ref{model2-obs-def}) we shift 
 \be
   \Omega_0+ \Omega_2~\rightarrow \Omega_0 + \Omega_2 + d_v \Omega_1
  \ee
with $\Omega_1$ being a $U(1)$ invariant form on $S^2$ we obtain 
   \bea
   && O_0~\rightarrow~ O_0 + \iota_v \Big (\frac{1}{2}\Omega_1~ \omega_{\mu\nu} \Psi^\mu \Psi^\nu  + \Omega_1 \e^a {\cal D}_a \Big )~,\nn\\
   && O_1~\rightarrow~ O_1 + \delta  \Big (\frac{1}{2}\Omega_1~ \omega_{\mu\nu} \Psi^\mu \Psi^\nu
    + \Omega_1 \e^a {\cal D}_a \Big ) + 
   \iota_v \Big  (\Omega_1 \omega_{\mu\nu} dX^\mu \Psi^\nu \Big )~,\\
   && O_2~\rightarrow~O_2 + \delta \Big (\frac{1}{2} \Omega_1 \omega_{\mu\nu} dX^\mu \Psi^\nu  \Big )
    + d \Big ( \frac{1}{2} \Omega_1 \omega_{\mu\nu} \Psi^\mu \Psi^\nu +
   \Omega_1 \e^a {\cal D}_a\Big )~.\nn
   \eea
Hence the observables get shifted only by $\delta$-exact terms. Thus the correlators of our observables depend only on the class in $H_{equiv}(S^2)$.  Next let us study the dependence on $(\e^a {\cal D}_a + \omega)$ within $H_{equiv}(M)$.  Consider the shift 
   \be
 \e^a {\cal D}_a + \omega ~\rightarrow~ \e^a {\cal D}_a + \omega + \e^a \iota_{k_a} \omega_1 + d\omega_1 + \e^a \iota_{k^a} \omega_3
   \ee
 assuming that $d\omega_3=0$ and $\omega_1$ is an invariant form (${\cal L}_{k_a} \omega_1=0$). This leads to
 \bea
 & O_0~\rightarrow~ & O_0 + \delta \Big ( \Omega_0 ~\omega_{1\mu} \Psi^\mu + 
  \frac{1}{6} \Omega_0 ~\omega_{3\rho\mu\nu} \Psi^\rho \Psi^\mu \Psi^\nu \Big) 
  - \iota_v \Big ( \Omega_0 ~\omega_{1\mu} dX^\mu + \frac{1}{2} \Omega_0~ \omega_{3\rho\mu\nu} dX^\rho \Psi^\mu \Psi^\nu \Big ) ~,\nn \\
  & O_1~\rightarrow~ & O_1 + d \Big ( \Omega_0 ~\omega_{1\mu}\Psi^\mu + 
   \frac{1}{6} \Omega_0~ \omega_{3\rho\mu\nu} \Psi^\rho \Psi^\mu \Psi^\nu \Big )
   - \delta \Big (\Omega_0~\omega_{1\mu} dX^\mu + \frac{1}{2} \Omega_0~
   \omega_{3\rho\mu\nu} dX^\rho \Psi^\mu \Psi^\nu \Big ) \nn\\
   && + \iota_v \Big ( \Omega_2~ \omega_{1\mu} \Psi^\mu + \frac{1}{6}
    \Omega_2~ \omega_{3\rho\mu\nu} \Psi^\rho \Psi^\mu \Psi^\nu -
     \frac{1}{2} \Omega_0 ~\omega_{3\rho\mu\nu} dX^\rho \wedge dX^\mu \Psi^\nu \Big )~,\\
     & O_2~\rightarrow~&O_2 - d \Big ( \Omega_0~\omega_{1\mu} dX^\mu 
      + \frac{1}{2} \Omega_0~\omega_{3\rho\mu\nu} dX^\rho  \Psi^\rho \Psi^\mu \Big) \nn \\
  &&    + \delta \Big (\Omega_2~\omega_{1\mu} \Psi^\mu 
      + \frac{1}{6} \Omega_2~ \omega_{3\rho\mu\nu} \Psi^\rho \Psi^\mu \Psi^\nu -
      \frac{1}{2} \Omega_0 ~\omega_{3\rho\mu\nu} dX^\rho \wedge dX^\mu \Psi^\nu \Big)~,\nn
 \eea
from which we conclude that the observables get shifted by $\delta$-exact term. Thus the correlators of our observables depend only on the corresponding class in $H_{equiv}(M)$. These arguments continue to hold for more general observables that will depend only on $H_{equiv}(S^2)$ and $H_{equiv}(M)$.

The partition function on $S^2$ corresponds to the following observable 
    \be\label{obs-equiv-both-action}
    2 \int\limits_{S^2} (O_0 + O_2) =  \int\limits_{S^2} (\Omega_0 + \Omega_2) (-\omega_{\mu\nu} dX^\mu \wedge dX^\nu + 2 \e^a
    {\cal D}_a + 
    \omega_{\mu\nu}  \Psi^\mu \Psi^\nu )~,
    \ee
where $\Omega_0+\Omega_2$ is given  by (\ref{2-form-sphere}) and we conveniently combined $O_0 + O_2$. As we discussed in the previous subsection $\Omega_0+\Omega_2$ can be decomposed into contributions of the north pole and of south pole. Thus we can argue the formal factorization in two contributions: one from the north hemisphere responsible for the dependence on $(b-i\omega)$ and one from  the south hemisphere responsible for the dependence on $(b+ i\omega)$.
       
We can also observe another property. From (\ref{model2-obs-rel}) we know that 
    \be
     d_v (O_0 + O_2) = \delta O_1~. 
    \ee
 We can set $\Psi=0$ and obtain
 \be
  d_v \Big ( (\Omega_0 + \Omega_2) (-\omega_{\mu\nu} dX^\mu \wedge dX^\nu + 2 \e^a
    {\cal D}_a ) \Big )=0
 \ee
provided that ${\cal L}_v X^\mu + \e^a k_a^\mu =0$ (formal equivariance of a map $X$). Thus the bosonic part of the observable can be evaluated on such equivariant maps
    \be
    \int\limits_{S^2} (\Omega_0 + \Omega_2) (-\omega_{\mu\nu} dX^\mu \wedge dX^\nu + 2 \e^a
    {\cal D}_a ) = 4\pi \e^a [{\cal D}_a (X_{np}) + 
   {\cal D}_a (X_{sp})]~,
    \ee
 where $X_{np}$, $X_{sp}$ are the values of  a map $X$ at the north/south poles
  which should coincide with the fixed points of $k_a$ due to the condition 
  ${\cal L}_v X^\mu + \e^a k_a^\mu =0$. Thus we should be careful with the implementing 
   condition ${\cal L}_v X^\mu + \e^a k_a^\mu =0$ directly since it tends to kill 
    all non-trivial topology related to non-trivial maps.
   If we restore the equivariant parameter $\varepsilon$ for $v$ on $S^2$ then in front of RHS we get the ratio $\e^a/\varepsilon$.

 \section{Localization locus revisited}\label{s:locus}
 
In this section we discuss the localization locus in some detail. At the moment we are unable to furnish a fully coherent mathematical description of the localization locus space and thus we limit ourselves to present some partial observations. Some of the PDEs which we discuss below have appeared previously in \cite{Nekrasov-Alex, Frenkel:2008vz} although in slightly different context.

\subsection{Model with equivariance on \texorpdfstring{$S^2$}{S2}}
 
In this subsection we analyze the model with equivariance on $S^2$, with the target space being any K\"ahler manifold. We use the cohomological description given in subsection \ref{ss:new-model-1} of the supersymmetric theory from section \ref{s:sigma-S2}. 
 
As follows from rewriting of the formulas (\ref{S2-exact-explicit}) and (\ref{new-eaxct-action}) the  supersymmetric action for the sigma model can be written as a sum of a BRST-exact term and the supersymmetric observable described in previous section
\begin{equation}
\begin{split}\label{loc-rewrting-action}
   &  S=\int\limits_{S^2} \Big ( dX^\mu \wedge \star dX^\nu g_{\mu\nu} + ...\Big ) \\ &     = \int \Big [ \delta (...)  + (\cos \theta + \Omega_2) (- \omega_{\mu\nu} dX^\mu \wedge dX^\nu + \omega_{\mu\nu} \Psi^\mu \Psi^\nu ) \Big ]~,
\end{split}
\end{equation}   
where the $\delta$-exact term is given by (\ref{S2-exact-explicit}). Due to standard arguments the path integral localizes on the solutions of the following PDEs
  \bea
 && (P^+ dX)^\mu=0~, \label{loc-eq-1A}\\
 && {\cal L}_v X^\mu =0~.\label{loc-eq-2A}
  \eea 
On these solutions we should evaluate the observables and calculate the corresponding one-loop determinant. Here we want to analyze these PDEs. It is impossible to find smooth solutions that satisfy all equations, hence we have to separate our analysis of the equations (\ref{loc-eq-1A}) and the equations (\ref{loc-eq-2A}). 
We first look in detail to the equations (\ref{loc-eq-1A}). Using the conventions from  subsection \ref{ss:projector} we have the following identity for our projector
  \be
  P^+ dX^i= \frac{z^2 d\bar{z} + dz}{1+|z|^4} \Big (\partial_z X^i + \bar{z}^2 \partial_{\bar{z}} X^i  \Big )
 \ee
which implies the following PDEs
 \be\label{eq-proj-explicit}
 \partial_z X^i + \bar{z}^2 \partial_{\bar{z}} X^i =0~,~~~~
 \partial_{\bar{z}} X^{\bar{i}} + z^2 \partial_{z} X^{\bar{i}}=0~. 
 \ee
These PDEs are not elliptic. At the south pole $z=0$ these equations reduce to $\partial_z X^i=0$ and at the north pole $z'=0$ the equations reduce to $\partial_{\bar{z}'} X^i=0$. As we move from the north pole to the south pole the equations (\ref{eq-proj-explicit}) smoothly transition between the two limits. We can solve the equations (\ref{eq-proj-explicit}) using the following trick. Let us define the map from $S^2$ to the disk $D$
   \be\label{def-map-S2D}
    i~:~S^2~\rightarrow~D
   \ee
with by the following explicit relations   
\be\label{def-map-S2D-extra}
  y = \frac{z}{1+|z|^2}~,~~~~~~ \bar{y} = \frac{\bar{z}}{1+|z|^2}~,
\ee
where $(y, \bar{y})$ are the coordinates on the disk $D = \{ |y|^2 \leq \frac{1}{4}\}$ and $(z,\bar{z})$ are the standard stereographic coordinates on $S^2$. Alternatively on another patch we can write
  \be
  y = \frac{\bar{z}'}{1+|z'|^2}~,~~~~~~ \bar{y} = \frac{z'}{1+|z'|^2}~,
 \ee
where we use the conventions from Appendix \ref{app:S2}. In polar coordinates this map is
  \be
   y = \frac{1}{2} \sin \theta~ e^{i\phi}~,~~~~~~~  \bar{y} = \frac{1}{2} \sin \theta ~e^{-i\phi}~.
  \ee
From these formulas we see that the map (\ref{def-map-S2D}) describes the double cover of $D$ by $S^2$ with the equator of $S^2$ ($|z|=1$) being mapped to the boundary of the disk $D$ ($|y|=\frac{1}{2}$). Using the map (\ref{def-map-S2D})  we can push forward the differential operators from $S^2$ to $D$. It is straightforward to derive  the following relations for the derivatives
   \be\label{deriv-S2D}
    \frac{\partial}{\partial z} = \frac{1}{(1+|z|^2)^2} \frac{\partial}{\partial y } - \frac{\bar{z}^2}{(1+|z|^2)^2} \frac{\partial }{\partial \bar{y}} =
     \frac{1}{4} \Big ( 1 + \sqrt{1-4 y\bar{y}} \Big )^2  \frac{\partial}{\partial y } - \bar{y}^2 \frac{\partial }{\partial \bar{y}}
   \ee
and its complex conjugate. Here we use the relation
     \be
       1- 4 y \bar{y} = \frac{(|z|^2-1)^2}{(1+ |z|^2)^2}~,
      \ee
and thus we had to make a choice for taking a square root.  The differential operator which appears in the equations (\ref{eq-proj-explicit}) has the following simple form on the disk
  \be
   (\partial_z + \bar{z}^2 \partial_{\bar{z}}) = \frac{1-|z|^2}{1+|z|^2} \partial_y    = \sqrt{1- 4 y \bar{y}} ~\partial_y~.
  \ee
Thus away from the equator $|z|=1$ the equations (\ref{eq-proj-explicit}) are mapped to $\partial_y X^i=0$ and $\partial_{\bar{y}} X^{\bar{i}}=0$ on the disk $D$.  In other words one can check explicitly that 
   \be\label{full-solution-DS}
   X^i(\bar{y})=  X^i \Big (\frac{\bar{z}}{1+|z|^2} \Big )~,~~~~~ X^{\bar{i}} (y) =X^{\bar{i}} \Big (\frac{z}{1+|z|^2} \Big )
    \ee
solve  the equations (\ref{eq-proj-explicit}). Hence we have reduced the non-elliptic system (\ref{eq-proj-explicit}) to anti-holomorphic maps from the disk $D$ to $M$ without specification of any boundary conditions. Let us  evaluate the bosonic part of the observable on the solutions of (\ref{eq-proj-explicit}). We start from the pull back of the symplectic form on the solutions of (\ref{eq-proj-explicit})
     \be
     \omega_{\mu\nu} (X)~ dX^\mu \wedge dX^\nu = (1-|z|^4)~ \partial_{z} X^{\bar{i}}
     \partial_{\bar{z}} X^j g_{\bar{i}j}(X)~  2i  dz \wedge d\bar{z}~.
   \ee
Assuming that $ 2i  dz \wedge d\bar{z}$ is positive, we see that the pull-back of $\omega$ takes a different sign on the upper and lower hemispheres and on the equator it becomes zero.  With our conventions ($z=0$, $\theta = \pi$ for the south pole and $z'=0$, $\theta=0$ for the north pole) we see that 
      \be
      \Big ( - \cos \theta~\omega_{\mu\nu} (X)~ dX^\mu \wedge dX^\nu \Big )~  \geq~ 0 ~.
      \ee
We can push to the disk $D$ also the second equation (\ref{loc-eq-2A}). The vector field $v = i (z \partial_z- \bar{z} \partial_{\bar{z}}) $ pushed to the disk $D$ has the following form \be
 i_*( v)=  i(y \partial_y - \bar{y} \partial_{\bar{y}})~,
\ee
hence the equation (\ref{loc-eq-2A}) becomes
   \be
   i (y \partial_y - \bar{y} \partial_{\bar{y}}) X^\mu =0~. 
   \ee
 If we impose this equation on the solutions (\ref{full-solution-DS}) we can see that the only smooth solutions are constant. However formally we may allow point like solutions, with holomorphic point like maps on the north pole and anti-holomorphic point like maps on the south pole.  We need better tools and better analytical control to enumerate such solutions and perform any reliable calculation. 
       
 One can suggest a different treatment of the problem. If the equations (\ref{eq-proj-explicit}) would give rise to some good moduli space then another equation  ${\cal L}_v X^\mu =0$ can be used to further localize on this moduli space with the $U(1)$-action coming from the rotation of $S^2$. In this picture the point-like solutions will be interpreted as fixed points under this $U(1)$-action in this good moduli space. The main problem is that the the equations (\ref{eq-proj-explicit}) can be converted to anti-holomorphic disk equations which do not give rise to a good moduli space unless appropriate boundary conditions are specified. For the (anti)holomorphic disks the good boundary conditions are when the boundary of  the disk is mapped to a Lagrangian submanifold. Let us suggest a possible logic which may lead to Lagrangian boundary conditions. If we take our localization argument (\ref{loc-rewrting-action}) and split the sphere in two hemispheres $S^2 = S^2_+ \cup S_-^2 $ with the boundary for each hemisphere  along  the equator $\gamma_{\rm eq}= \partial S_+^2 = - \partial S_-^2$ then we can try to run the localization argument for each hemisphere.  For the localization to work on each hemisphere we have to impose appropriate boundary conditions on the equator and in the path integral we sum over all allowed boundary conditions. We can argue for the correct boundary conditions in the following fashion. From our previous discussion about the observables we know that  $\delta O_2 = dO_1$. The integral over the north hemisphere $S_+^2$ of $O_2$ is not BRST invariant since  
     \be\label{BRST-equator}
      \delta \int\limits_{S_+^2} O_2 = \int\limits_{\gamma_{\rm eq}} O_1 =
     \int\limits_{\gamma_{\rm eq}} ( \cos \theta + f(\theta)) ~\omega_{\mu\nu} (X) dX^\mu \Psi^\nu
     \ee
unless we require that $X$ maps the equator to Lagrangian submanifold. Here in the observable $O_1$ we have inserted $( \cos \theta + f(\theta))$ following our discussion around equation (\ref{class-2form}). We want that our observable defined over the hemisphere still depends only on the class in the equivariant cohomology of $S^2$ and not on a concrete representative. In this case $O_1$ restricted to $\gamma_{\rm eq}$ is identically zero and the integral of $O_2$ over $S_+^2$ is good observable. We can apply the same argument for the other hemisphere $S_-^2$. If we accept this logic then the equations (\ref{eq-proj-explicit}) get supplemented by Lagrangian boundary conditions and we can reduce the problem to anti-holomorphic disks with Lagrangian boundary conditions. Thus in the path integral we have to further localize in each moduli space and then sum up over all allowed boundary conditions. At the moment this is a rather speculative logic and one needs to study problem further in order to perform some non-trivial checks. 
   
\subsection{Model with target space equivariance}
   
In this subsection we study the localization locus when the target space admits some torus action compatible with the K\"ahler structure. The corresponding cohomological field theory was described in subsection \ref{ss:model-2}. As follows from the BRST exact term (\ref{model2-exact-action}) the localization locus is given by the following PDEs
    \bea
  &&  (P^+ (dX + \e^a k_a A))^\mu  = 0~,\label{loc-eq1-mod2}\\
  && \iota_v \Big ( dX^\mu + \epsilon^a k_a^\mu A \Big ) =  {\cal L}_v X^\mu + \e^a k^\mu_a (X)=0~,\label{loc-eq2-mod2}
  \eea
where we  introduce the singular flat connection
    \be
     A= - \frac{i dz}{2 z} + \frac{i d\bar{z}}{2\bar{z}}
    \ee
such that 
    \be
     \iota_v A_v =1~. 
    \ee
As before we analyze these two equations separately. We start from the equations (\ref{loc-eq1-mod2}) which in complex coordinates can be written explicitly as follows
  \be
  \partial_z X^i + \bar{z}^2 \partial_{\bar{z}} X^i = \frac{i}{2} (z^{-1} - \bar{z}) \e^a k_a^i~,
  \ee
and its complex conjugate. Here $(z,\bar{z})$ are the stereographic coordinates on $S^2$. Following the discussion from the previous subsection we can introduce the map $i:S^2\rightarrow D$ with $(y,\bar{y})$ defined in (\ref{def-map-S2D-extra}). Using (\ref{def-map-S2D-extra}) and (\ref{deriv-S2D}) the above equations become
   \be
   (1 - |z|^2) \Big ( y \partial_y X^i -\frac{i}{2} \e^a k_a^i \Big )=0
     \ee
on the disk $D$. Thus we obtain the following PDEs on for the interior of $D$
  \be\label{hol-eq}
   y \partial_y X^i = \frac{i}{2} \e^a k_a^i~,~~~~~~
    \bar{y} \partial_{\bar{y}} X^{\bar{i}} = -\frac{i}{2} \e^a k_a^{\bar{i}}~.
  \ee
Since we deal with the Hamiltonian action $k_a^\mu = \omega^{\mu\nu} \partial_\nu {\cal D}_a$ on a K\"ahler manifold we can rewrite these equations as follows
   \be\label{loceq-nice-form}
    y \partial_y X^i = \frac{1}{2} g^{i\bar{j}} \partial_{\bar{j}} (\e^a {\cal D}_a)~,
    ~~~~~~~~
     \bar{y} \partial_{\bar{y}} X^{\bar{i}} = \frac{1}{2} g^{\bar{i}j} \partial_{j} (\e^a {\cal D}_a)~.
   \ee
Our goal is to understand how to solve these equations. Let us look at a very simple example when we consider the maps from the disk to $\mathbb{C}$. In complex coordinates we have $X$ and $\bar{X}$ with $k^\mu\partial_\mu = i (X\partial_X - \bar{X} \partial_{\bar{X}})$. Hence the above equations are
  \be\label{elementary-example}
   y \partial_y X = - \frac{1}{2} \e X~,~~~~~ 
   \bar{y} \partial_{\bar{y}} \bar{X} = - \frac{1}{2} \e \bar{X}~.
  \ee
These equations have the following simple solutions
  \be
    X \sim |y|^{-\frac{1}{2}\e} \bar{y}^n~,~~~~~~
    \bar{X} \sim |y|^{-\frac{1}{2}\e} y^n~,
  \ee
where $n$ is a non-negative integer and $\e <0$ for the solutions to be smooth. Actually one can see that the  solutions of the equations (\ref{loceq-nice-form}) have the following form
    \be\label{space-solutions-gen}
     X^i (\bar{y},  |y|)~,~~~~~
       X^{\bar{i}} (y, |y|) ~,
    \ee
 since 
 \be
     y \partial_y  X^i (\bar{y},  |y|) =  \frac{1}{2} |y|  \frac{\partial   X^i (\bar{y},  |y|)}{\partial |y|} = \frac{1}{2} g^{i\bar{j}}  \partial_{\bar{j}} (\e^a {\cal D}_a) ~,
\ee
where the derivative with respect to $|y|$ acts only on the second argument in $X^i(y,|y|)$. Hence we have anti-holomorphic disks twisted by an additional radial dependence $|y|$ which is controlled by the gradient flow with the Morse function given by the moment map $\e^a {\cal D}_a$. For example, if we look only on the solutions of the form $X^\mu(|y|)$ then  
    \be
     \frac{d X^\mu}{d t} = g^{\mu\nu} \partial_\nu (\e^a {\cal D}_a)
    \ee
with $|y|=e^t$, hence we deal with the gradient flows between the fixed points. Since the disk has a finite size $|y| \leq \frac{1}{2}$, $t\in (-\infty, - \log 2)$.  Thus the solutions of (\ref{loceq-nice-form}) are some mixture of  anti-holomorphic disks and gradient flows for the Morse function given by the moment map $\e^a {\cal D}_a$. As we discussed previously for anti-holomorphic disks to give rise to a nice moduli space  we have to impose the Lagrangian boundary conditions on the boundary of $D$. The argument presented around the equation (\ref{BRST-equator}) can be repeated here and most likely the good moduli space of (\ref{loceq-nice-form}) would require imposing  Lagrangian boundary conditions  for $D$. Moreover in this fully equivariant story the Lagrangian submanifolds may be required to preserve some toric symmetry (as may be argued from (\ref{loc-eq2-mod2})). 
  
Provided that we have a good moduli space we can understand the equation 
 \be  
 {\cal L}_v X^\mu + \e^a k_a^\mu = 0
 \ee
as a further localization on the moduli space. Formally this equation will tell us that the fixed points of $S^2$ are mapped to the fixed points of $k_a$ on $M$.  We cannot find any smooth solutions satisfying all equations. Rewritten in $(y,\bar{y})$ coordinates we have  
   \be
      (y \partial_y - \bar{y} \partial_{\bar{y}}) X^\mu = i \e^a k_a^\mu~. \label{real-eq}
      \ee
and there are no smooth single valued solutions for both equations (\ref{real-eq}) and (\ref{loceq-nice-form}). To illustrate this let us look at the example of maps from $D$ to $\mathbb{C}$ (see above) which satisfy both (\ref{elementary-example}) and the equations
   \be
     (y \partial_y -   \bar{y} \partial_{\bar{y}} ) X = - \e X~,~~~ (y \partial_y -   \bar{y} \partial_{\bar{y}} ) \bar{X} =  \e \bar{X}
   \ee
where we have assumed $\e$ to be real. It is  hard to find solutions for all equations. The only solution we find
   \be
    X = y^{-\e/2} \bar{y}^{\e/2}~,~~~~  \bar{X} = \bar{y}^{-\e/2} y^{\e/2}~,
       \ee
that is not single valued.   

\section{1-loop calculation}\label{s:1-loop}
 
In this section we perform the localization calculation around the constant maps. We start with the model without target space equivariance from subsection \ref{ss:new-model-1} and we reintroduce the equivariant parameter $\varepsilon$ in front of $v$ in the transformations (\ref{transf-model-1}) but we keep the canonical kinetic term as in \eqref{new-eaxct-action}. In the localization locus we have a distinguished subspace of constant maps $dX=0$ which solves both equations (\ref{loc-eq-1A}) and (\ref{loc-eq-2A}). The moduli space of constant maps is identified with the target space $M$.  We denote by $X^\mu_0$ a point on target K\"ahler manifold $M$ and by $\Psi^\mu_{0}$ the corresponding fermionic zero modes which have the interpretation of $dX_0^\mu$. We will localize around such constant solutions and calculate explicitly the one-loop determinant. For the contribution of other topological sectors we will make a conjecture in the next section. The localization answer for the observable \eqref{observ-non-eq} for such maps has the following form
\begin{equation}
\begin{split}\label{one-loop-formal}
  Z_{S^2}\left(\varepsilon , t ,\tb \right) = \int\limits_M d^d X_0~ d^d \Psi_0~ Z^{(const)}_{\mathrm{1-loop}}(\varepsilon, X_0, \Psi_0) \exp\left(i\varepsilon^{-1} \left(t^a-\bar{t}^a\right) \omega_a (X_0)_{\mu\nu} \Psi^\mu_0 \Psi^\nu_0 \right) \\
  = \int\limits_M  Z^{(const)}_{\mathrm{1-loop}}(\varepsilon) \exp\left(i\varepsilon^{-1}\left(t^a-\bar{t}^a\right) \omega_a \right) 
  \end{split}
\end{equation}
where we will use standard notations for differential forms and the 1-loop contribution $Z^{(const)}_{\mathrm{1-loop}}$ is understood as a differential form. Here in the observable \eqref{observ-non-eq} we have to have an extra factor $\varepsilon^{-1}$ since we require a canonical kinetic term. Hence the observable should contain the term $\cos \theta~ \omega_{\mu\nu}dX^\mu \wedge dX^\nu$ but in the transformations we should have $\varepsilon$ in front of ${\cal L}_v$. 
Here $\omega_a$ is a basis in $H^{2}(M)$ and  $t^a$, $\bar{t}^a$ are the complexified coordinates such that 
 \be
  (-i \omega + b) = \sum_{a=1}^{\dim H^2} t^a \omega_a~.
 \ee
Hence in the exponent of \eqref{one-loop-formal} we have the evaluation of the observable \eqref{observ-non-eq} on the constant maps with a canonical kinetic term.  
 
The  one-loop contribution $Z^{(const)}_{\mathrm{1-loop}}(\varepsilon)$ can be obtained by the study of the linearized supersymmetry transformations \eqref{transf-model-1} around the constant maps. Let us expand our variables $X$ and $\Psi$ around the constants maps and related zero modes
\begin{equation}\label{def-fluct-both}
    X^\mu = X^\mu_0 + \varDelta X^\mu \qquad \Psi^\mu = \Psi_0^\mu+ \varDelta\tilde{\Psi}^\mu
\end{equation}
where $\varDelta X^\mu$ and $\varDelta\tilde{\Psi}^\mu$ are the bosonic and fermionic fluctuations over which we have to integrate. The problem is that while the bosonic fluctuation $\varDelta X^\mu$ transforms tensorially (with respect to the diffeomorphisms of $M$) the fermionic fluctuation $\varDelta\tilde{\Psi}^\mu$ does not transform tensorially and thus we have a problem in defining the path integral measure. The standard way to fix it is to use the Levi-Civita connection and redefine 
     \be\label{def-new-fluct}
     \varDelta\tilde{\Psi}^\mu = \varDelta \Psi^\mu -\Gamma^\mu_{~\nu \rho} \Psi_0^\nu \varDelta X^\rho~.
     \ee
Now $\varDelta \Psi^\mu$ transform tensorially and thus the path integral measure is well-defined. We have to track down the corresponding supersymmetry transformations for the fluctuations. We plug \eqref{def-fluct-both} and \eqref{def-new-fluct} into \eqref{transf-model-1}, and asume the following transformations for the zero modes $\delta X_0=\Psi_0$, $\delta \Psi_0=0$~. We obtain the following linearized transformations for the fluctuations $ \varDelta X^\mu$ and $\varDelta{\Psi}^\mu$ 
\begin{equation}
    \begin{split}\label{eq:susyfluct}
        & \delta\varDelta X^\mu=\varDelta \Psi^\mu -\Gamma^\mu_{~\nu \rho} \Psi_0^\nu \varDelta X^\rho~,\\
        &\delta \varDelta \Psi^\mu = \varepsilon\L_v \varDelta X^\mu + \dfrac{1}{2}R^\mu_{~ \nu \lambda } \Psi_0^\nu \Psi_0^\lambda~ \varDelta X^\sigma- \Gamma^\mu_{~\nu \rho} \Psi_0^\nu ~\varDelta \Psi^\rho~.
    \end{split}
    \end{equation}
It is convenient to define the covariant version of the transformations  $\delta_\nabla = \delta + \Gamma^\mu_{\nu \rho}\Psi^\nu_0$ hence 
\begin{equation}
\begin{split}
&\delta_\nabla \varDelta X^\mu = \varDelta \Psi^\mu \\
&\delta_\nabla \varDelta \Psi^\mu = \varepsilon \L_v \varDelta X^\mu + \dfrac{1}{2}R^\mu_{~\sigma \nu \lambda} \Psi_0^\nu \Psi_0^\lambda \varDelta X^\sigma
 =  \varepsilon \L_v \varDelta X^\mu + (R)^\mu_{~\sigma} \varDelta X^\sigma
\end{split}
\end{equation}
where we use the short hand convention $(R)^\mu_{~\sigma} = \dfrac{1}{2}R^\mu_{~\sigma \nu \lambda} \Psi_0^\nu \Psi_0^\lambda$. The fields $\chi$ and $H$ do not have any zero modes for the constant maps and thus we treat them as fluctuations. The original transformations \eqref{transf-model-1} are linear in the $\chi$ and $H$ fields and we should further linearize them in the fluctuations $\varDelta X^\mu$ and $\varDelta{\Psi}^\mu$. The final result can be written as follows
\begin{equation}
\begin{split}\label{lin-tr-Hchi}
  &  \delta_\nabla \chi^\mu =H^\mu ~,\\
  &  \delta_\nabla H^\mu = \varepsilon \mathcal{L}_v \chi^\mu+ (R)^\mu_{~\nu} \chi^\nu~, \\
\end{split}
\end{equation}
where we use again the short hand conventions $(R)^\mu_{~\nu}$ and $\mathcal{L}_v = \mathcal{L}^\Gamma_v$ at the linearized level in the fluctuations. 

Using the linearized transformations \eqref{eq:susyfluct}, \eqref{lin-tr-Hchi} and applying standard arguments (see e.g. the review \cite{Pestun:2016jze} or the more detailed exposition in \cite{Festuccia:2019akm}) we can derive the one-loop contribution. It is given by the following superdeterminant
\begin{equation}
Z^{(const)}_{\mathrm{1-loop}}(\varepsilon)= \operatorname{sdet}^{1/2}\left( \delta_\nabla^2 \right)=\operatorname{sdet}^{1/2}\left( \varepsilon\mathcal{L}_v+ \hat{R} \right)= \left(\dfrac{\operatorname{det}_{\chi^\mu}\left( \varepsilon\mathcal{L}_v+ \hat{R} \right)}{\operatorname{det}_{\Delta X^\mu}\left( \varepsilon \mathcal{L}_v+ \hat{R} \right)} \right)^{1/2}~,
\end{equation}
where the denominator comes from \eqref{eq:susyfluct} and   the numerator comes from \eqref{lin-tr-Hchi}.  Here $\hat{R}$ is the Lie algebra valued curvature form $(R)^\mu_{~\nu}$ and the determinant is assumed both over the infinite set of modes and the Lie algebra action on $TM$.  Recall that  $\Delta X$ are zero forms on $S^2$ with values in the tangent bundle of $M$
\begin{equation}\label{space-delta-X}
\Delta X^\mu \in \Omega^{0}\left(S^2, X^*(TM) \right)
\end{equation}
while the odd field $\chi$ is in the subspace of 
one-forms on $S^2$ valued in the tangent bundle $M$ with respect to the $P^{+}$ projector
\begin{equation}\label{space-chi}
\chi^\mu \in \Omega^{1+}\left(S^2, X^*(TM) \right)~.
\end{equation}
(see the discussion around \eqref{def-omega1TM} for more explanations). 
We stress that in \eqref{space-delta-X} and \eqref{space-chi} we deal with the linearized spaces around $X_0$, therefore below we regard $\Omega^{0}\left(S^2, X^*(TM_{X_0}) \right)$    $\Omega^{1+}\left(S^2, X^*(TM_{X_0}) \right)$ as linear spaces (we suppress the expansion point $X_0$ in our notation). Hence  the one-loop answer is written more properly as follows
\begin{equation}
    Z^{(const)}_{\mathrm{1-loop}}(\varepsilon) =  \left(\dfrac{\operatorname{det}_{\Omega^{1+}\left(S^2, X^*(TM) \right)}\left( \varepsilon\mathcal{L}_v+ \hat{R} \right)}{\operatorname{det}_{\Omega^0\left(S^2, X^*(TM) \right)}\left( \varepsilon\mathcal{L}_v+ \hat{R} \right)} \right)^{1/2}~.
\end{equation}
This ratio of determinants can be calculated in different ways and it is related to the one-loop contribution of the chiral field on $S^2$. We present below an explicit calculation of this determinant ratio stressing the relevant regularization issues.
Instead of diagonalizing explicitly he relevant operator, we exploit the fact that there are many cancellations. For that we use the operator $P^+d$ that connects the two spaces
\begin{equation}
 P^+d : \quad \Omega^{0}\left(S^2, X^*(TM) \right) \xrightarrow{P^{+}d} \Omega^{1+}\left(S^2, X^*(TM) \right)
\end{equation}
and commutes with the operator which appears in the one-loop expression 
\begin{equation}
[\mathcal{L}_v+ \hat{R},P^{+}d]=0~.
\end{equation}
There are cancellations  between common eigenfunctions so that the answer reduces to the kernel and cokernel
\begin{equation}
 \left(\dfrac{\operatorname{det}_{\Omega^{1+}\left(S^2, X^*(TM) \right)}\left( \varepsilon\mathcal{L}_v+ \hat{R} \right)}{\operatorname{det}_{\Omega^{0}\left(S^2, X^*(TM) \right)}\left( \varepsilon\mathcal{L}_v+ \hat{R} \right)} \right)^{1/2}= \left(\dfrac{\operatorname{det}_{\operatorname{coker} P^+d}\left(\varepsilon \mathcal{L}_v+ \hat{R} \right)}{\operatorname{det}_{\operatorname{ker} P^+d}\left( \varepsilon\mathcal{L}_v+ \hat{R} \right)} \right)^{1/2}
\end{equation}
We now address in detail how to find the kernel and the cokernel, since they admit explicit descriptions on $S^2$. First we make use of the target space complex structure and separate the determinant into that over the holomorphic and anti-holomorphic parts of the tangent bundle $T^{(1,0)}M$ and $T^{(0,1)}M$. 
We use stereographic coordinates $(z, \zb)$ on $S^2$; by changing the coordinates to $(z', \zb')$ we can check that everything is well-defined on $S^2$.
On the holomorphic part the kernel is defined by the equation
\begin{equation}
P^{+}d \,    f^{i}(z,\zb) =0 \Longleftrightarrow
\left( 
\partial_z + \bar{z}^2  \partial_{\zb}
 \right) 
f^i(z,\zb) =0 
\end{equation}
which, as discussed above is solved by any function of the form:
\begin{equation}
f^{i}(z,\zb)=f^{i}\left( \dfrac{\zb}{1+|z|^2} \right)
\end{equation}
Hence the kernel is spanned by basis functions (or every $i$)
\begin{equation}
f^{i,(m)}(z,\zb)\sim  \left( \dfrac{\zb}{1+|z|^2} \right)^m ,  \ m\geq 0
\end{equation}
which is diagonal with respect to ${\cal L}_v$. The cokernel is defined by the one-forms $\chi^i$ which are not in the image of $P^+d$ operator. Equivalently we can write 
\begin{equation}
d^\dagger \chi^i =0~,~~~~~P^- \chi=0
\end{equation}
where we use Hodge decomposition and the second condition makes sure that $\chi^i$ is in the correct space.
These two conditions are solved by (for every $i$)
\begin{equation}
\chi^{i,(m)} \sim \frac{1}{(1+|z|^2)^{2}}\left(\frac{z}{(1+|z|^2)}\right)^{m}\left(d z+z^{2} d \bar{z}\right) \, , \ m\geq 0
\end{equation}
For the anti-holomorphic components (for every $\bar{i}$) we get the complex conjugate answer 
\begin{equation}
\begin{split}
\operatorname{ker}  P^{+}d & = \left\{  \left(\frac{z}{(1+|z|^2)}\right)^{m}  ,  \ m\geq 0 \right\}
\\
\operatorname{coker}  P^{+}d & = \left\{ \frac{1}{(1+|z|^2)^{2}}\left(\frac{\zb}{(1+ |z|^2)}\right)^{m}\left( \zb^2 d z+ d \bar{z}\right) \, ,  \ m\geq 0 \right\}
\end{split}
\end{equation}
On $S^2$ these basis element diagonalize the Lie derivative operator ${\cal L}_v$
\begin{equation}
  \mathcal{L}_v f^{i,(m)}= - i m  f^{i,(m)}~,  \qquad \mathcal{L}_v \chi^{i,(m)}= i
   (m+1)   \chi^{i,(m)}~.
\end{equation}
We also diagonalize the curvature form $\hat{R}$ and denote the Chern roots by $r_i$. Finally the ratio of determinants is given by
 the following expression
\begin{equation}\label{infinite-prod-1loop}
Z^{(const)}_{\mathrm{1-loop}}(\varepsilon)=\left( \dfrac{\prod\limits_{i=1}^{\dim_{\mathbb{C}} M} \prod\limits_{m=1}^{\infty} ( \varepsilon m+  r_i)^2}{ \prod\limits_{i=1}^{\dim_{\mathbb{C}} M} \prod\limits_{m=1}^{\infty} ( \varepsilon m-  r_i)^2} \right)^{1/2}~,
\end{equation}
where we have excluded the constant mode ($m=0$) of the $\varDelta X^\mu$ fluctuations since we expand around the constant maps. Using zeta-function regularization the infinite products can be rewritten in terms of  Gamma functions as follows
\begin{equation}
     \left. \left(\prod_{m=1}^\infty \dfrac{1}{\varepsilon m + x} \right) \right|_{reg}  =
\frac{\varepsilon}{2 \pi} \Gamma\left( 1+ x/\varepsilon\right)  \varepsilon^{ x / \varepsilon} ~.
\end{equation}
The resulting answer as a function of the Chern roots can be presented in terms of the characteristic class called the Gamma class \cite{Galkin:2014laa,libgober1998chern}  as 
\begin{equation}\label{ration-gamma}
  Z^{(const)}_{\mathrm{1-loop}}(\epsilon)  =\varepsilon^{\frac{2 c_1(M)}{\varepsilon}} \dfrac{\Hat{\Gamma}_M\left(\varepsilon\right)}{\bar{\Hat{\Gamma}_M}(\varepsilon)}~,
\end{equation}
where $c_1(M)$ is the first Chern class of $M$. The Gamma class is a multiplicative characteristic class which can be expressed through the  Chern characters $\operatorname{ch}_k\left(TM\right)$ 
\begin{equation}
    \hat{\Gamma}_M (\varepsilon)=\det\Gamma\left(1+\dfrac{\hat{R}}{\varepsilon}\right)=\exp \left(-\gamma_{Eu}\dfrac{c_{1}(M)}{\varepsilon}+\sum_{k \geqslant 2}(-1)^{k}(k-1) ! \zeta(k) \dfrac{\operatorname{ch}_{k}(T M)}{\varepsilon^k}\right)~.
\end{equation}
The conjugate is given by $\bar{\Hat{\Gamma}_M}(\varepsilon)=\det\Gamma\left(1-\frac{\hat{R}}{\varepsilon}\right)$.
Finally substituting \eqref{ration-gamma} into \eqref{one-loop-formal} we can write down the full answer for 
 the contribution of the constant maps
\begin{equation}\label{1loop-full} 
    Z_{S^2}\left(\varepsilon , t ,\tb \right) = \int\limits_M  \   \varepsilon^{\frac{2 c_1(M)}{\varepsilon}} \dfrac{\Hat{\Gamma}_M\left(\varepsilon\right)}{\bar{\Hat{\Gamma}_M}(\varepsilon)} \exp\left(i\varepsilon^{-1} \left(t^a-\bar{t}^a\right) \omega_a \right)~.
\end{equation}
This answer is written for any K\"ahler manifold $M$. As it stands this integral is well-defined for compact $M$ and it depends only on the cohomology class of $\hat{R}$ and the cohomology class of the complexified K\"ahler form $\omega$. 
 
If we assume that $M$ is a Calabi-Yau (CY) manifold ($c_{1}(X)=0$) then the dependence of \eqref{1loop-full} on the equivariant parameter $\varepsilon$ is analytical
\begin{equation}\label{CYanswer}
     Z_{S^2}\left(\varepsilon , t ,\tb \right) = \int\limits_M  \  \dfrac{\Hat{\Gamma}_M\left(\varepsilon\right)}{\bar{\Hat{\Gamma}_M}(\varepsilon)} \exp\left(i\varepsilon^{-1}\left(t^a-\bar{t}^a\right) \omega_a \right)
\end{equation}
This answer agrees with the results presented in \cite{Halverson:2013qca,Hori:2013ika}. In \cite{Halverson:2013qca} 
the answer \eqref{CYanswer} which encodes the perturbative corrections to the K\"ahler potential was argued from mirror symmetry considerations and some other consistency checks (see also  \cite{Morrison:2016bps} for a review). In \cite{Hori:2013ika} the above answer was discussed in the context of a GLSM localization calculation and its geometrical meaning (we will comment more on this case below). Here our goal was to obtain this answer directly from the calculation within the non-linear sigma model without assuming any specific geometrical restrictions on $M$. Due to degree considerations upon the expansion the answer \eqref{CYanswer} has the following overall dependence 
\be
  Z_{S^2}\left(\varepsilon , t ,\tb \right) \sim \varepsilon^{-\frac{\dim M}{2}}
\ee
in agreement with the anomaly considerations in  \cite{Gomis:2015yaa} (keeping in mind that $\varepsilon$ is proportional to $R^{-1}_{S^2}$).

Let us write down some explicit formulas for CY manifolds in different dimensions. If we look at a CY-threefold then expanding \eqref{CYanswer} to the appropriate order we get
\begin{equation}\label{const-CY3}
     Z_{S^2}\left(\varepsilon , t ,\tb \right) =   - \dfrac{2\zeta(3)}{\varepsilon^3} \int\limits_M  c_3(M) - \dfrac{i}{3!\varepsilon^3}\sum_{a,b,c} (t^a -\tb^a) (t^b -\tb^b) (t^c -\tb^c) \int\limits_M  \omega_a \wedge \omega_b \wedge \omega_c ~,
\end{equation}
where the first integral on the RHS is just the Euler number of $M$. The $\zeta(3)$ term as a perturbative correction was initially obtained in \cite{Grisaru:1986dk} by explicit evaluation of loop integrals in the $\mathcal{N}=(2,2)$ sigma model and later via mirror symmetry in \cite{Candelas:1990rm}.
For higher dimensional CY the perturbative contributions mix with the observable, for example for a CY-fourfold we obtain
\begin{equation}
\begin{split}
  &   Z_{S^2}\left(\varepsilon , t ,\tb \right)  =   - \dfrac{2i\zeta(3)}{\varepsilon^4}\sum_{a} (t^a -\tb^a) \int\limits_M c_3 (M) \wedge \omega_a \\
   &  + \dfrac{1}{4!\varepsilon^4}\sum_{a,b,c,d} (t^a -\tb^a) (t^b -\tb^b) (t^c -\tb^c)(t^d -\tb^d)  \int\limits_M  \omega_a \wedge \omega_b \wedge \omega_c \wedge \omega_d~.
     \end{split}
\end{equation}
Notice, that a potential $\zeta(4)$ perturbative contribution is absent. This is in accordance with the absence of the five loop correction in the sigma model \cite{Grisaru:1986wj}. In fact in the ratio of gamma functions only terms with odd zeta-values survive, and the transcendental weight is related to the loop order at which corrections to the sigma model appear. This pattern of seemingly complicated transcendental loop corrections summing up into a simple expression indicates once again, why the study of the transcendental structure of loop integrals \cite{Broadhurst:1996kc,Bonisch:2021yfw,Brown:2015fyf} is important.

Let us comment on the non-analytical dependence in \eqref{1loop-full} for a general K\"ahler case.  In \eqref{1loop-full} we can collect the terms with $c_1(M)$ as follows
\begin{equation}
    \exp\left(\frac{ 2 \,  c_1(M) }{\varepsilon} \left( \log \varepsilon - \gamma_{Eu} \right)\right)~. 
\end{equation}
For example, if we look at a six dimensional K\"ahler manifold  then in \eqref{const-CY3} the perturbative term on RHS should be replaced as follows for the non CY case
\begin{equation}\label{WithScale}
     - 2 \zeta(3) \int_M c_3(M) \longrightarrow - 4\zeta(3) \int_M \operatorname{ch}_3(M) + \dfrac{4\left(\log \varepsilon - \gamma_{Eu}\right)^3}{3} \int_M \left( c_1(M) \right)^3~. 
\end{equation}
We believe that one can relate $ \varepsilon$ to the $UV$ renormalization scale (there will be some ratio of $\varepsilon$ and the UV scale). Since the model is no longer conformal, a dependence on such parameter is expected. In particular we conjecture that if one restores the scale dependence in \cite{Grisaru:1986dk,Grisaru:1986wj}, one would obtain exactly the expression \eqref{WithScale}. We think that a similar statement can be made for K\"ahler manifold of other dimensions and the corrections analyzed in  \cite{Grisaru:1986dk,Grisaru:1986wj} can be better understood  in cohomological terms. 
 
 \subsection{One loop with target space equivariance}
 
We analyzed the constant map contribution to the partition function for the model with just $S^2$ equivariance. The answer \eqref{1loop-full} is given in terms of an integral over certain de Rham cohomology classes, this integral is well-defined for a compact K\"ahler target space manifold. For non-compact examples, it may require an additional regularization. If a target space admits isometries then target space equivariance may serve as natural way to regularize the non-compact answer. Here we briefly sketch the derivation for the model with the target space equivariance. The main idea is that the result is the same as in \eqref{1loop-full} but now all classes are understood as equivariant classes and the integral can be localized and written as a sum over fixed point contributions. 

Let us analyze the contribution of constant maps in the model with target space equivariance. In this case a constant map will provide a solution to the localization equations \eqref{loc-eq1-mod2} and \eqref{loc-eq2-mod2} if $k_a (X_0)=0$. Hence we have a finite number of constant maps which map $S^2$ to the fixed points of our torus action on $M$. Let us concentrate on one given fixed point $X_0^\mu$ (we assume that all fixed points are isolated) and analyze the linearized supersymmetry \eqref{def-trans-model2} around this point. The main difference is that in this case there are no fermionic zero modes $\Psi_0$ and thus we will regard $\Psi$ as fluctuations, together with the $\chi$ and $H$ fields. Up to linear order in the fluctuations we obtain the following transformations
\begin{equation}
    \begin{split}
        & \delta \Delta X^\mu = \Psi^\mu~,\\
     & \delta  \Psi^\mu  = \varepsilon \mathcal{L}_v \Delta X^\mu + \e^a \partial_\nu k^\mu_a (X_0) \Delta X^\nu~,\\
        &  \delta \chi^\mu = H^\mu~,\\
     & \delta H^\mu = \varepsilon {\cal L}_v \chi^\mu + \e^a \partial_\nu k^\mu_a(X_0) \chi^\nu~. 
    \end{split}
\end{equation}
 At the fixed point $X_0$ we have $\partial_\nu k^\mu_a (X_0)= \nabla_{\nu} k^\mu_a(X_0)$. Due to the isometry property this is an anti-symmetric matrix acting on $TM_{X_0}$ (for short below we denote this matrix as $\partial k (X_0)$). Next we have to calculate the determinants and this goes through exactly in the same way as we discussed earlier. Eventually we arrive to the expression \eqref{infinite-prod-1loop} where under $i r_i$ we understand the eigenvalues of $\partial k (X_0)$. Moreover we now have to keep the mode $m=0$ for the $\Delta X$ fluctuations. Finally we can summarize the one-loop contributions for a given fixed point $X_0$ as follows
\begin{equation}\label{1loop-equiv-one} 
    Z_{\mathrm{1-loop}}(\varepsilon,\epsilon^a, X_0)= e^{2 \frac{\log \varepsilon}{\varepsilon} \e^a \Tr (\partial k_a (X_0))}
\frac{1}{\sqrt{\det (\e^a  \partial k_a (X_0))}}  \dfrac{ \det \Gamma\left (1+ \frac{\e^a  \partial k_a (X_0)}{\varepsilon} \right )}{\det \Gamma \left (1-\frac{\e^a \partial k_a (X_0)}{\varepsilon} \right ) } ~,
\end{equation}
where the determinants are understood on $TM_{X_0}$ and the additional determinant in front of Gamma functions comes from the $m=0$ contribution. It is important to stress that~\eqref{1loop-equiv-one} is a complicated expression in the equivariant parameters and it does not terminate if we expand it. In order to write the full answer we have to sum over all fixed points $X_0$ and evaluate the observable \eqref{obs-equiv-both-action} at the constant maps $X_0$. Hence the full contribution can be written as follows
    \be\label{full-equiv-sum}
 Z_{S^2} (\varepsilon, \e^a ) = \sum\limits_{X_0}~ Z_{\mathrm{1-loop}}(\varepsilon,\epsilon^a, X_0)~ e^{8\pi \frac{\e^a}{\varepsilon} {\cal D}_a(X_0)}~, 
    \ee
 where the factor $4\pi$ comes from the evaluation of the volume of $S^2$ in our conventions. The additional factor $\varepsilon^{-1}$ in the exponent is related to the canonical normalization of the kinetic term and thus to fixing the normalization of the observable \eqref{obs-equiv-both-action} with all parameters turned on. The answer \eqref{full-equiv-sum} can be written as an integral over of the appropriate equivariant forms as follows
 \be\label{full-eq-integral}
  Z_{S^2} (\varepsilon, \e^a ) = \int\limits_M ~ \frac{\det\Gamma (1+\dfrac{\hat{R}_{\rm equiv}}{\varepsilon} )}{\det\Gamma(1-\dfrac{\hat{R}_{\rm equiv}}{\varepsilon} )} ~e^{\frac{8\pi}{\varepsilon} \omega_{\rm equiv}+ 2 \frac{\log \varepsilon}{\varepsilon}
   \Tr (\hat{R}_{\rm equiv})}~.
\ee
Hence this is the equivariant extension of the previously discussed answer \eqref{1loop-full}. Here the equivariant extensions are 
 defined as $\omega_{equiv} = \omega + \e^a {\cal D}_a$ and $\hat{R}_{\rm equiv} = \hat{R} + \e^a \nabla k_a$ where 
  we have used the following identity 
\be
 \nabla_\alpha \nabla_\sigma k^\rho_a + R^\rho_{~\sigma\gamma\alpha} k^\gamma_a=0~,
\ee
which follows from the standard properties of the curvature tensor and the fact that the $k_a$ are isometries of the corresponding metric. By applying the Berline-Vergne-Atiyah-Bott localization theorem to \eqref{full-eq-integral} we reproduce the answer  \eqref{full-equiv-sum} with \eqref{1loop-equiv-one}. 
  
If we impose the CY condition\footnote{We need to set to zero  the equivariant first Chern class $\Tr (\hat{R}_{\rm equiv})$. For this beside the CY condition we also need to require that the isometries preserve the CY stucture, i.e. $\sum_a \e^a=0$.}  the answer \eqref{full-equiv-sum} has the following functional dependence
    \be 
    Z_{S^2} (\varepsilon, \e^a )  = \varepsilon^{-\frac{\dim M}{2}} F\left (\frac{\e^a}{\varepsilon}\right )~,
  \ee
hence it agrees with the anomaly considerations in  \cite{Gomis:2015yaa}. Please observe that the natural normalization of the target space equivariant parameters is $\e^a \varepsilon^{-1}$ since the flat connection \eqref{def-connection} has $\varepsilon^{-1}$ in front to keep the property $\iota_v A=1$. 
Finally let us remark about the relation to the localization result for GLSM. Let us restrict to the toric CY manifolds which are obtained by the K\"ahler quotient $\mathbb{C}^N//(U(1))^r$ then the integral over the equivariant characteristic classes \eqref{full-eq-integral} can be written in terms of $r$-dimensional contour integral (see \cite{Nekrasov:2021ked} for an explanation). This integral can be identified with the concrete pole contribution (within the zero flux sector) in the full answer for the GLSM model of the corresponding toric CY manifold.

\section{Summary and full answer}\label{s:summary}

 In this paper we have concentrated on the formulation on $S^2$ of supersymmetric $N=(2,2)$ non-linear sigma models with a K\"ahler target manifold. We described these supersymmetric theories on $S^2$ and provided a reformulation in terms of cohomological theory similar to the A-model. 
 Unlike the A-model here we introduce a new notion of 2D self-duality defined on one forms $\Omega^1(\Sigma, X^*(TM))$ that uses the existence of $U(1)$ vector field on $S^2$. We also considered the model with target space equivariance, corresponding to a supersymmetric sigma model coupled to a supersymmetric background gauge multiplet. We have analyzed the observables in the model and we have presented a discussion of the localization locus and of the 1-loop calculation around constant maps.
     
 Let us present a conjectured structure for the full result. If we look at the model with a general K\"ahler target space manifold then we have argued that the localization locus is given by holomorphic disks which presumably should end on Lagrangian submanifolds (although we cannot derive this directly within the present framework). Hence we can conjecture that the full answer should be written schematically as sum over all Lagrangian submanifolds
    \be\label{general-answer}
    Z_{S^2} = \sum\limits_{L} Z_{L}
    \ee
where $Z_L$ is the theory associated to the moduli space of holomorphic disks ending on a given $L$. This theory is not the A-model, the counting should be done differently and the relevant cohomology class is related to the ratio of two Gamma classes (as we saw in the case of the constant maps). The conjectured answer is hard to check in such generality, but some explicit checks may be done for simple examples of target spaces. Another related question is how the structures related to A-model appear in the present context. For example, it is not clear to us how the quantum cohomology (ring structure of observables) appear in the present context. Many standard arguments from the A-model cannot be applied directly here, for example the local observables are stuck at  fixed points etc. 
   
 If we move to the case when the model has target space equivariance then the situation is better. For example, if we assume that the target space manifold is a toric CY manifold then we expect that the answer for $Z_{S^2}$ should be given in terms of a GLSM localization calculation (certain sum over fluxes of contour integrals). The main challenge is to understand how GLSM answer encodes the counting of disks at the level of the non-linear sigma model. It would be natural to expect that the formula \eqref{general-answer} holds but now we have to sum over a specific class of invariant Lagrangians. It would be nice to perform some simple enumerative calculations for the disks and to understand how to extract them from the GLSM answer. This can also help us to understand better the localization locus we discussed. 
 
 Let us make a final comment, the relation of the presently discussed non-linear sigma model with all equivariant parameters $\varepsilon$ and $\e^a$ to the A-model is not as simple as it may appear at first. It is not so easy to extract the non-equivariant answer and the claim that we deal with gluing of $A$ and $\bar{A}$ models is not that straightforward. We think that the role of the equivariant parameters should be studied better and one should pay more attention to different expansions in the parameters of the model. We hope to come back to these issues elsewhere.

\bigskip
{\bf Acknowledgements:} We are very grateful to Tobias Ekholm for illuminating discussions.  
The work of Guido Festuccia is supported by Vetenskapsr\aa{}det under grant 2018-05572. The work of Maxim Zabzine  is supported by the grant  ``Geometry and Physics"  from the Knut and Alice Wallenberg foundation. The work of Victor Mishnyakov and Victor Alexeev was partly funded by RFBR and MOST, project number 21-52-52004 (V.M.) and RFBR grant 20-01-00644 (V.A, V.M.).

 \appendix
 
\section{Notations}\label{app:conv}
\label{sec:notations}

The Euclidean flat-space metric is $\delta_{A B}$, $A,B=1,2$ and the Levi-Civita symbol $\epsilon^{A B}$ is normalized to $\epsilon^{12}=1$. We use complex coordinates $z= x^1 + i x^2, \bz = x^2 - i x^2$, so that $\delta_{z\bz}= \half$, $\delta_{zz}=\delta_{\bz\bz}=0$ and $\epsilon^{z\bz}= -2 i$.  

We denote Weyl spinors with $\pm$ indices so that $\psi_-$ and $\psi_+$ have spin $\half$ and $-\half$, respectively, under ${\rm Spin}(2) \cong U(1)$. We also use Dirac spinors
\be\label{Diracspin}
\psi = (\psi_\alpha)= \begin{pmatrix} \psi_- \cr \psi_+\end{pmatrix}~.
\ee
The two-dimensional gamma matrices are ${(\gamma^{A})_{\alpha}}^\beta={ (-\sigma^1, -\sigma^2)_{\alpha}}^\beta$ when $A$ runs over $1, 2$, and $\gamma^3 = \sigma^3$, with $\sigma^A$ the Pauli matrices. They satisfy $\gamma^A \gamma^B = \delta^{A B}+i \epsilon^{A B}\gamma^3$ and $\{ \gamma^3,\gamma^A\}=0$.
In complex coordinates, we have
\be\label{gammazzb}
\gamma_z= \begin{pmatrix}0 & 0 \cr -1 & 0 \end{pmatrix}~, \qquad \gamma_\bz= \begin{pmatrix}0 & -1 \cr 0 & 0 \end{pmatrix}~.
\ee
Dirac indices are raised and lowered with the epsilon symbols $\epsilon^{\alpha\beta}, \epsilon_{\alpha\beta}$ and are contracted from upper-left to lower-right so that
\be\label{expprod}
\psi \chi = \psi_+ \chi_- -\psi_- \chi_-~, \qquad
\psi \gamma^3\chi = \psi_+ \chi_- +\psi_- \chi_-~.
\ee

The supersymmetry covariant derivatives read
\bea
\label{defDpii}
& D_+ = {\d \over \d \theta^+} - 2 i \tilde \theta^+ \partial_{\b z}~, \qquad \widetilde D_+ = - {\d \over \d {\tilde \theta}^+} + 2 i  \theta^+ \partial_{\bar z}~,  \cr
&D_- = {\d \over \d \theta^-} + 2 i \tilde \theta^- \partial_z~, \qquad \widetilde D_- = - {\d \over \d {\tilde \theta}^-} - 2 i  \theta^- \partial_z~,
\eea
they satisfy 
$\{D_- , \widetilde D_- \} = - 4 i \d_z$ and $\{D_+ , \widetilde D_+ \} =  4 i \d_{\bar z}$.

\section{Summary of $S^2$ geometry}\label{app:S2}

In this appendix we summarize our conventions on $S^2$ geometry. The unit sphere $S^2$ is defined in 
$\mathbb{R}^3$ as 
  \be
   x_1^2 + x_2^2 + x_3^2 = 1~.
  \ee
We define the stereographic projection of $S^2$ minus point $(0,0,1)$ (we refer to this point as the north pole) into the plane $x_3=0$ 
  \be
  \label{deftheta}
   z= \frac{x_1 + i x_2}{1-x_3} = \cot \frac{\theta}{2}~e^{i\phi}~,
  \ee
where $(\theta, \phi)$ are the standard spherical coordinates with the values $\phi \in [0, 2\phi)$, $\theta \in [0, \pi]$. Analogously we define the stereographic projection of $S^2$ minus point $(0,0,-1)$ (we refer to this point as south pole) into the plane $x_3$
    \be
    z'= \frac{x_1-ix_2}{1+x_3} = \tan \frac{\theta}{2}~ e^{-i\phi}~.
    \ee
Using these two stereographic projections we see that for $S^2$ minus north and south poles 
 \be
  z= \frac{1}{z'}
 \ee
and this diffeormophism identifies $S^2$ with the projective space $\mathbb{CP}^1$. Throughtout the paper we use two systems of coordinates for $S^2$, either $(z, \bar{z})$ (with $(z', \bar{z}')$ on another patch) or the spherical coordinates $(\theta, \phi)$ and the following relation will be useful
     \be
    \cos \theta = \frac{|z|^2 -1}{1+|z|^2}= \frac{1-|z'|^2}{1+|z'|^2}
     \ee
with the appropriate regions of validity.  In our conventions the north pole of $S^2$ corresponds to $\theta= 0, z'=0$ and the south pole to $\theta=\pi, z=0$.
   
The sphere $S^2$ is equipped with the Fubini-Study metric 
   \be
  \frac{4 dz d\bar{z}}{(1+ |z|^2)^2} = d\theta^2 + \sin^2 \theta d\phi^2
 \ee
and with the Fubini-Study K\"ahler form
   \be
  \Omega_2 = \frac{2 i ~dz\wedge d\bar{z}}{(1+ |z|^2)^2} =  \sin \theta~ d\phi \wedge d\theta~,\label{def-Omega2}
 \ee
which also defines the volume form. The sphere $S^2$ admits $U(1)$-action which is realized by shifting the angle $\phi$ (or phase of $z$). The corresponding vector field is defined as follows
    \be
    v = \partial_\phi = i (z\partial_z - \bar{z} \partial_{\bar{z}})~,
    \ee
and we define the dual 1-form
    \be
     \kappa = g(v) = \frac{2i (zd\bar{z}-\bar{z} dz)}{(1+|z|^2)^2} = \sin^2 \theta~d\phi~,
    \ee
such that 
     \be
     ||v||^2 = \iota_v \kappa = g(v,v) = \frac{4|z|^2}{(1+|z|^2)^2}~. 
    \ee
This $U(1)$-action is the Hamiltonian since 
   \be
    (d + \iota_v) (\Omega_2 + \Omega_0)=0
   \ee
with the Hamiltonian function $\Omega_0 = \cos \theta$. 
 
Sometimes we consider the sphere $S^2$ of radius $R_{S^2}$. In this case we use the following  form  for the Fubini-Study metric
 \be
  \frac{4R_{S^2}^2 dz d\bar{z}}{(1+ |z|^2)^2} =  R_{S^2}^2(d\theta^2 + \sin^2 \theta d\phi^2)~,
 \ee
where we use the conventions with $z$ being dimensionless coordinate. By appropriate rescaling of $z$ one can switch to other convention with the dimensionful coordinates.

\section{Cohomological complex}
\label{seccoh}
 
Here we present the explicit map between the ${\cal N}=(2,2)$ chiral and anti-chiral superfield components and the cohomological fields introduced in sections~\ref{ss:new-model-1} and~\ref{ss:model-2}~.

In order to define the cohomological fields we will use the Killing spinors $\zeta$ and $\tilde \zeta$. When the supersymmetric background admits more than one $\zeta$ or $\tilde \zeta $ we need to single out one of each to construct the cohomological variables. 

From the Killing spinors we can form the spinor bilinears:
\bea
&&v^m= -2\zeta\gamma^m\tilde \zeta~,\qquad s= \tilde \zeta(1-\gamma_3)\zeta~,\qquad \tilde s= \tilde \zeta(1+\gamma_3)\zeta~,\cr 
&&\tau_m=\zeta \gamma_m \zeta~,\qquad \tilde\tau_m= \tilde \zeta \gamma_m\tilde \zeta~.
\eea
The vector field $v$ is Killing and the scalars $s$ and $\tilde s$ are constant along $v$.

We require $s+\tilde s$ and $s^2+\tilde s^2$ to be smooth positive real functions. We can then write down the following projectors acting on one forms (we remind the reader that we denote with $\kappa$ the one form with components $\kappa_m=v_m$)
\bea
&&P^{+}={s+\tilde s\over 2(s^2+\tilde s^2)}\left((s+\tilde s)1\!\!1+ i (\tilde s-s)\star- {1\over s+\tilde s} \kappa\wedge \iota_v\right)~,\cr
&&\widetilde P^{+}={s+\tilde s\over 2(s^2+\tilde s^2)}\left((s+\tilde s)1\!\!1- i (\tilde s-s)\star- {1\over s+\tilde s} \kappa\wedge \iota_v\right)~.
\eea
Using these definitions it follows that $P^+\tau=\tau$ while $\widetilde P^+\tilde \tau=\tilde \tau$~.

Specifically we are interested in the round two sphere and we select the spinors~\eqref{2dspin} with $A=\tilde A=1~, B=\tilde B=0$~. With this choice the projectors $P^+$ is:
 \be
   P^+ = \frac{1}{2(1+|z|^4)} \Big ( 1 + |z|^4 + i (1-|z|^4) \star 
    + 2 (z^2 d\bar{z}\wedge \iota_{\partial_z} + \bar{z}^2 dz \wedge \iota_{\partial_{\bar{z}}}) \Big )~,
 \ee
which can be compared with~\eqref{projzzb} when acting on one forms with the values in $X^*(T^{1,0}M)$. 

We start by considering the case of chiral and anti-chiral multiplets that are not coupled to a background gauge field. The cohomological variables for the chiral multiplet (see section~\ref{chirachir}) are defined as follows:
\bea
\label{cohdef}
X^i&=&X^i\cr
\Psi^i &=&  \sqrt{2} \zeta\psi^i~,\cr
\chi^i_m&=& {\sqrt{2}\,\tau_m \over s+\tilde s}\, \tilde\zeta \psi^i~,\cr
h^i_m&=& \tau_m F^i+i{s^2+\tilde s^2\over s+\tilde s} \left(P^+\partial X^i\right)_m~.
\eea 
Note that we have $P^+\chi^i=\chi^i$ and $P^+h^i=h^i$~. The inverse map is given by
\bea
\psi^i&=&-\sqrt 2 \left({2\over s^2+\tilde s^2}\tilde\tau^m \chi^i_m  \zeta +{1\over s+\tilde s} \Psi^i \tilde \zeta \right)~,\\
\label{inverseF}
F^i&=&-{2\over s^2+\tilde s^2} \tilde \tau^m h^i_m+{2i\over s+\tilde s}\tilde \tau^m \partial_m X^i~.
\eea
The supersymmetry variation $\delta=\delta_\zeta+\delta_{\tilde \zeta}$ acts on the cohomological variables as:
\bea
\label{susyco}
\delta X^i &= & \Psi^i~,\cr
\delta \Psi^i &=& i v^m \partial_m X^i~,\cr
\delta \chi^i_m &=& h^i_m~,\cr
\delta h^i_m &=&  i ({\cal L}_v \chi^i)_m~.
\eea 
These coincide with~\eqref{cohomj} up to redefining $\delta$ when acting on Grassmann odd variables with an extra factor of $-i$.

Next we consider the case where there is an Abelian isometry generated by a holomorphic Killing vector field $k^i$. We can then couple the chiral fields to a supersymmetric background gauge field. The definition of the cohomological variables is then
\bea
\label{cohdefiso}
X^i&=& X^i\cr
\Psi^i &=&  \sqrt{2} \zeta\psi^i~,\cr
\chi^i_m&=& {\sqrt{2}\, \over s+\tilde s}\,\tau_m\, \tilde\zeta \psi^i~,\cr
h^i_m&=& \tau_m F^i+i{s^2+\tilde s^2\over s+\tilde s} \left(P^+D X^i\right)_m+(\sigma-\tilde \sigma){\tilde \zeta \gamma^3\tilde \zeta\over s+\tilde s}  \tau_m k^i~.
\eea 
where $D_m X^i= \partial_m X^i-k^i$. The inverse map is
\bea
\psi^i&=&-\sqrt 2 \left({2\over s^2+\tilde s^2}\tilde\tau^m \chi^i_m  \zeta +{1\over s+\tilde s} \Psi^i \tilde \zeta \right)~,\\
\label{inverseFga}
F^i&=&-{2\over s^2+\tilde s^2} \tilde \tau^m h^i_m+{2i\over s+\tilde s}\tilde \tau^m D_m X^i- (\sigma-\tilde\sigma){\tilde \zeta \gamma^3\tilde \zeta\over s +\tilde s}k^i~.
\eea
Supersymmetry acts on the cohomological variables as:
\bea
\label{susycoga}
\delta X^i &= & \Psi^i~,\cr
\delta \Psi^i &=& i v^m \partial_m X^i -i (v^n A_n-i (s\sigma+\tilde s \tilde \sigma ) ) k^i~,\cr
\delta \chi^i_m &=& h^i_m~,\cr
\delta h^i_m &=&  i ({\cal L}_v \chi^i)_m-i (v^n A_n-i (s\sigma+\tilde s \tilde \sigma ) )\partial_j k^i \chi^j_m~.
\eea 
For the Killing spinors we selected on the round sphere and with the choice of background gauge field as in~\eqref{bckgh} we have that $$v^m A_m -i (s\sigma+\tilde s \tilde \sigma)={u\over R_{S^2}}$$ where $u$ is a constant complex parameter. Hence the susy transformations simplify to
\bea
\label{susycois}
\delta X^i &= & \Psi^i~,\cr
\delta \Psi^i &=& i v^m \partial_m X^i- i{u\over R_{S^2}} k^i~,\cr
\delta \chi_m^i &=& h_m^i
\cr
\delta h_m^i &=&   i ({\cal L}_v \chi^i)_m-i {u\over R_{S^2}}\partial_j k^i \chi^j_m~.
\eea 
Comparing with~\eqref{cohomequiv} we identify the target-space equivariant parameter to be $\epsilon=-{u\over R_{S^2}}$~.

With these definitions the transformations of the cohomological variables are independent of the sigma model target space geometry. However the fields $h^i$ do not transform as tensors under holomorphic coordinate changes. As we reviewed in section~\ref{s:cohomological} we can define fields $H^i=h^i +\Gamma^i_{j k} \Psi^j \chi^k$ that do transform tensorially. 

Similarly we can consider an anti-chiral field of $R$-charge 0. The corresponding $\tilde\chi^i$ and $\tilde h^i$ fields are then in the image of $\widetilde P^+$~:
\bea
\label{cohdefac}
 X^{\bar i}&=&\widetilde X^{\bar i}&\cr
 \Psi^{\bar i} &=&  -\sqrt{2} \tilde \zeta\tilde \psi^{\bar i}~,\cr
 \chi^{\bar i}_m&=& -{\sqrt{2}\, \over s+\tilde s}\, \tilde \tau_m\, \zeta \tilde \psi^{\bar i}~,\cr
 h^{\bar i}_m&=& \tilde \tau_m \tilde F^{\bar i}+i{s^2+\tilde s^2\over s+\tilde s} \left(\widetilde P^+D \widetilde X^{\bar i}\right)_m+(\sigma-\tilde \sigma){ \zeta \gamma^3 \zeta\over s+\tilde s}  \tilde \tau_m k^{\bar i}~.
\eea
Next we look at the action obtained from the Lagrangian density~\eqref{dchiral}. Using the Killing spinors we selected on the round sphere and with the choice of background gauge field as in~\eqref{bckgh} the action can be rewritten in terms of the  cohomological variables. This results in the observable ${\cal O}_2$ introduced in~\eqref{model2-obs-def} plus various $\delta$-exact terms:
\begin{align}
\label{actionrew}
  &S= {1\over 2} \int \Big ( -\cos(\theta)\, \omega_{\mu\nu} dX^\mu \wedge dX^\nu + 2\Omega_2 \e^a {\cal D}_a  - i \, \Omega_2\,\omega_{\mu\nu} \Psi^\mu \Psi^\nu \Big)+\cr
  &\quad+ \int d^2x \sqrt{g_{S^2}} g_{\mu \nu} \delta \left[ {2 \over 1+\cos(\theta)^2 }\chi^\mu_m  H^{\nu m} -{2i} \chi^\mu_m D^m X^{\nu} +\right.\cr
  &\quad\qquad \left.-{i \over 2} \Psi^\mu\iota_v D X^{\nu}-i {u f \over 2 R_{S^2}} \Psi^\mu k^{\nu}\right]~.
\end{align}
Here the angle $\theta$ is the latitude on $S^2$ as in~\eqref{deftheta} while $\Omega_2$ is defined in~\eqref{def-Omega2}. The indices $\mu,\nu$ run over $i,\bar i$. The arbitrary function $f(z\bar z)$ that specifies the background gauge multiplet configuration~\eqref{bckgh} appears only in $Q$-exact terms.

\section{Squashing}\label{squash}

We can squash the two sphere maintaining a $U(1)$ isometry by multiplying the radius $R_{S^2}$ by a positive function $c(z\bar z)$. By rescaling $R_{S^2}$ the values of the function $c$ at the two poles can be taken to be $w^2=c(\infty)=c(0)^{-1}$~.  The maximum number of Killing spinors satisfying~\eqref{Killing} is reduced to two. For instance 

 \bea
 \label{2dspinsq}
 &&\zeta_*=\begin{pmatrix}
    \zeta_-  \\ \zeta_+
  \end{pmatrix}= \frac{\sqrt{c(z\bar z)}}{\sqrt{(1+ z \b z)}}\begin{pmatrix}
     w  \\
 {i\over w  } z
  \end{pmatrix}~,\\
 && \tilde \zeta_*=\begin{pmatrix}
   \tilde \zeta_-  \\ \tilde \zeta_+
  \end{pmatrix}= \frac{\sqrt{c(z\bar z)}}{\sqrt{(1+ z \b z)}}\begin{pmatrix}
 {i\over w }  \b z  \\
    w
  \end{pmatrix}~. \nonumber
 \eea
satisfy the Killing spinor equations for vanishing background $U(1)_R$ connection and
 \be
 {\cal H}= \frac{i w^2}{R_{S^2}} \frac{ c- (1+z\bar z) c'}{c^2}~,\qquad  \tilde {\cal H}= \frac{i}{w^2 R_{S^2}}
 \frac{c+(1+z \bar z) c'}{c^2}~.
 \ee  
The spinors bilinears~\eqref{bildef}, built from the two Killing spinors above are
\be
 v= {i\over R_{S^2}}(z\partial_z -\bar z \partial_{\bar z })~,\quad s={z \bar z\over 1+ z \bar z}{c\over w^2}~,\quad \tilde s ={1\over 1+z \bar z}w^2 c~.
\ee
Near the $z=0$ pole the spinors~\eqref{2dspinsq} and the  supercharge corresponding to $\delta_Q=\delta_\zeta+\delta_{\tilde \zeta}$ approach those corresponding to the $\bar A$ topological twist. Similarly near the $z=\infty$ pole the supercharge approaches that corresponding to the $A$ topological twist. 

\subsection{background gauge field}

Here we present the supersymmetric background gauge field configurations that preserve both supercharges on the squashed sphere. Introducing a function $f(z \bar z)$ we can set
\bea\label{bckght}
&& \quad A=-{i u\over 2}\left(1- {w^2+ w^{-2} z \bar z\over 1+z\bar z} c f\right)\left({dz\over z}-{d\bar z \over \bar z} \right)~,\cr
&& \sigma=i {u f\over R_{S^2}} ~,\quad \tilde \sigma=i  {u f\over R_{S^2}}~,\quad D= {u f\over c R_{S^2}^2}{w^2+w^{-2}\over 2}-u{(1+ z \bar z)(w^2-w^{-2} z \bar z  )\over 2 c^2  R_{S^2}^2 } (c f)'~,
\eea
where $u$ is a complex constant.
The anticommutator of $\delta_{\tilde \zeta}$ and $\delta_\zeta$ reduces to
\be
 \{\delta_{\tilde \zeta}, \delta_\zeta \}\phi = i {\cal L}_v\phi+ \frac{1}{2R_{S^2}} \left(r+2 q u\right)\phi~.
\ee
Hence neither the squashing profile $c(z \bar z)$ nor the function $f(z \bar z)$ appear in the superalgebra. 

\providecommand{\href}[2]{#2}\begingroup\raggedright

\bibliographystyle{utphys}
\bibliography{exoticA-model}{}

\endgroup

\end{document}